\documentclass[twocolumn]{aastex631}

\usepackage{booktabs}
\usepackage{multirow}
\usepackage{array}
\usepackage{hyperref}
\newcolumntype{H}{>{\setbox0=\hbox\bgroup}c<{\egroup}@{}}
\usepackage{efbox,graphicx}
\usepackage{stackengine,xcolor}
\efboxsetup{linecolor=black,linewidth=1pt}
\usepackage[caption=false]{subfig}
\definecolor{nsgreen}{rgb}{0.1,0.5,0.1}

\usepackage{lineno}
\usepackage{longtable}
\usepackage{lineno}
\shorttitle{FRAMEx. V. Multi-frequency VLBA observation of AGNs}
\shortauthors{Shuvo et al.}
\graphicspath{{./}{figures/}}

\begin{document}

\title{FRAMEx. V. Radio Spectral Shape at Central Sub-parsec Region of AGNs}

\correspondingauthor{Onic Islam Shuvo}
\email{oshuvo@gmu.edu}

\author[0000-0003-4727-2209]{Onic I. Shuvo}
\affiliation{Department of Physics, University of Maryland Baltimore County, 1000 Hilltop Circle, Baltimore, MD 21250, USA}
\affiliation{U.S. Naval Observatory, 3450 Massachusetts Ave NW, Washington, DC 20392-5420, USA}

\author[0000-0002-4146-1618]{Megan C. Johnson}
\affiliation{U.S. Naval Observatory, 3450 Massachusetts Ave NW, Washington, DC 20392-5420, USA}

\author[0000-0002-4902-8077]{Nathan J. Secrest}
\affiliation{U.S. Naval Observatory, 3450 Massachusetts Ave NW, Washington, DC 20392-5420, USA}

\author[0000-0002-8818-9009]{Mario Gliozzi}
\affiliation{Department of Physics and Astronomy, George Mason University, MS3F3, 4400 University Drive, Fairfax, VA 22030, USA}


\author[0000-0002-8736-2463]{Phillip J. Cigan}
\affiliation{U.S. Naval Observatory, 3450 Massachusetts Ave NW, Washington, DC 20392-5420, USA}     

\author[0000-0002-3365-8875]{Travis C. Fischer}
\affiliation{AURA for ESA, Space Telescope Science Institute, 3700 San Martin Drive, Baltimore, MD 21218, USA}

\author[0000-0001-9149-6707]{Alexander J. Van Der Horst}
\affiliation{Department of Physics, The George Washington University, 725 21st St. NW, Washington, DC 20052, USA}




\begin{abstract}

We present results from the Very Long Baseline Array (VLBA) multi-frequency  ($1.6, 4.4, 8.6, 22$~GHz),  high-sensitivity  ($\sim$25 $\mu$Jy~beam$^{-1}$), sub-parsec scale ($<$1 pc) observations and Spectral Energy Distributions (SEDs) for a sample of 12 local active galactic nuclei (AGNs), a subset from our previous volume-complete sample with hard X-ray (14--195~keV) luminosities above $10^{42}$~erg~s$^{-1}$, out to a distance of 40~Mpc. All 12 of the sources presented here were detected in the C (4.4 GHz) and X (8.6 GHz) bands, 75\% in the L band(1.6 GHz), and 50\% in the K band (22 GHz).  Most sources showed compact, resolved/slightly resolved, central sub-parsec scale radio morphology, except a few with extended outflow-like features. A couple of sources have an additional component that may indicate the presence of a dual-core, single or double-sided jet or a more intricate feature, such as radio emission resulting from interaction with nearby ISM. The spectral slopes are mostly GHz-peaked or curved, with a few showing steep, flat, or inverted spectra. We found that in the sub-parsec scale, the GHz-peaked spectra belong to the low-accreting, radio-loud AGNs with a tendency to produce strong outflows, possibly small-scale jet, and/or have a coronal origin. In contrast, flat/inverted spectra suggest compact radio emission from highly-accreting AGNs' central region, possibly associated with radio-quiet AGNs producing winds/shocks or nuclear star formation in the vicinity of black holes.
\end{abstract}



\section{Introduction} \label{sec:intro}

An active galactic nucleus (AGN) generates enormous power. It radiates throughout the electromagnetic spectrum from radio to gamma rays due to the accretion of matter onto the Supermassive Black Hole (SMBH) residing at its center. To monitor and observe such a powerful and luminous heart of the active galaxies, the U.S. Naval Observatory (USNO), in collaboration with a few other institutions, is conducting the Fundamental Reference AGN Monitoring Experiment, or FRAMEx~\citep{Dorland_2020jsrs.conf..165D}. Using ground and space-based telescopes, FRAMEx observed a sample of nearby AGNs in the X-ray and radio wavelengths at multiple epochs to understand and characterize their physical properties. 

In FRAMEx I,~\citet[hereafter, paper~I]{Fischer_2021ApJ...906...88F} tested the ``Fundamental Plane of Black Hole Activity'' \citep[e.g.,][hereafter the FP]{Merloni_2003MNRAS.345.1057M,Gultekin_2009ApJ...706..404G}, the apparent correlation between radio emission (6 GHz), X-ray emission (2-10 keV), and black hole mass ($M_\mathrm{BH}$), using a simultaneous Swift X-ray Telescope (XRT) and Very Long Baseline Array (VLBA) radio observations of 25 local AGNs. If it held, it would support the unification of AGN accretion physics in conflict with the purported radio-loud (RL)/radio-quiet(RQ) dichotomy~\citep{hutchings_1989ApJ...342..660H,Miler_1990MNRAS.244..207M,antonucci_1993ARA&A..31..473A}. But it was found that the FP breaks down when probing the innermost regions of AGNs in milliarcsecond (sub-parsec) scales. In contrast, the extended extra-nuclear radio emission (10s to 100s of parsec) measured from archival Jansky Very Large Array (VLA) observations is responsible for the apparent FP correlation. A few other FRAMEx results provided intriguing motivation for a natural extension of our earlier observations. For example, only three out of nine sources were detected in a follow-up campaign~\citep[hereafter, paper~III]{Shuvo_2022} with higher observing sensitivity ($\sim$ 10$\mu$Jy) of initially undetected sources (from Paper I) at 6 GHz despite reaching noise levels $\sim 2$ times below the predicted rms from the VLA detections ($\sim$ 50$\mu$Jy), thus calling into question whether radio-quiet AGNs may, in fact, be radio silent at VLBA spatial scales. In addition, a drop in apparent radio luminosity was observed for NGC 2992 in our first simultaneous X-ray and radio variability campaign~\citep[hereafter, Paper~II]{fernandez_2022ApJ...927...18F}. NGC 2992 was identified clearly in the initial 6 GHz radio survey, but during six epochs of monitoring this AGN every 28 days, we observed the luminosity fall below the detection limit during the fourth session by at least a factor of three and then reappear in the final two epochs. Also, the possibly discrepant extended structure of NGC 1068 across multiple epochs in the C-band raises exciting questions about the variability of the source structure. All these results underscore the importance of obtaining quasi-simultaneous, multi-wavelength radio observations to understand the physical processes that affect the apparent luminosity, variability, and morphologies of AGNs. 

The limited accessible bandwidth of the VLBA C-band (6 GHz) observations alone precluded the determination of the radio spectral index $\alpha$ (in this paper, $\alpha$ is defined as $S_\mathrm{\nu}\propto \nu^\mathrm{+\alpha}$), which can be used to distinguish between competing radio emission mechanisms, e.g., synchrotron versus thermal bremsstrahlung (free-free). Observations from L ($\sim$1.6 GHz) to K-band ($\sim$22 GHz) at VLBA resolutions span the necessary frequency range to shed light on the physical processes at work in these AGNs. By utilizing source spectral indices and resolving the compact cores at higher radio frequencies, we can gain a deeper understanding of the physical nature of these systems.

Radio emission in a radio-loud AGN is primarily caused by synchrotron emission. This occurs when highly collimated plasma jets are accelerated away from the nucleus at relativistic speeds. 
On the other hand, in a radio-quiet AGN, the radio emission could be produced by a wide range of physical mechanisms, such as star formation, accretion disc winds, coronal disc emission, or low-power jets, without any significant contribution from the powerful relativistic luminous jets~\citep{panessa_2019NatAs...3..387P}. Following these different origins of radio emission, Paper III showed from a sub-parsec scale set of VLBA observations of radio-quiet AGNs that relativistic particles accelerated in shocks and winds or low-luminosity outflows may be a possible source of weak radio emission (and consequent low detection rate). We demonstrated that the non-detections might result from synchrotron self-absorption at 6 cm in the radio core, similar to what has been observed in stellar-mass black holes in X-ray Binaries (XRBs) while transitioning from the radiatively inefficient state to a radiatively efficient state~\citep[and references therein]{Coriat_2011MNRAS.414..677C}. A well-sampled radio spectral energy distribution can help us determine the true origin of radio emission and disentangle contributions from multiple potential components.

To achieve this goal, we present results from a high-resolution,
multi-wavelength radio continuum emission study of 12 detected AGNs from our previous C-band campaigns (Papers I, II, and III). We obtained VLBA observations in the L ($\sim$1.6 GHz), C ($\sim$4.4 GHz), X ($\sim$8.6 GHz), and K ($\sim$22 GHz) radio bands with homogeneous observing setups (details are discussed in Section~\ref{subsection: VLBA}), instrument systematics, and integration to a similar sensitivity level providing a consistent baseline from which meaningful comparisons can be made. The sample selection, multi-wavelength VLBA radio observations, and data calibration details are discussed in Section~\ref{sec:Methodology}. We present our results and discussion about the origin of the radio emission for these nearest hard X-ray selected AGNs in Section~\ref{sec:res} and Section~\ref{sec:dis}, respectively. 

\begin{deluxetable*}{lrrlcccHHH} 
\tablecaption{FRAMEx Multi-frequency Sample}
\tablehead{\colhead{Target} & \colhead{R.A.\ (ICRS)} & \colhead{Decl.\ (ICRS)}  & \colhead{Type} & \colhead{Redshift} & \colhead{Distance} & \colhead{log($M_\mathrm{BH}$)} \\
[-0.3cm]
& \colhead{(deg)}  & \colhead{(deg)}    &  & & \colhead{(Mpc)} & \colhead{[$M_{\sun}$]} }
\startdata
NGC 1052 & 40.2699938 & $-$8.25576443   & Sy2   & 0.0050 & 21.5 & 8.67 & 40.15 & 0.11 & 2.2$\times10^{-3}$\\
NGC 1068 & 40.6696215 & $-$0.01329436   & Sy2   & 0.0038 & 16.3 & 6.93 & 41.34 & 1.74 & 3.5$\times10^{-2}$\\
NGC 2110 & 88.0474036 & $-$7.45625207  & Sy2   & 0.0078 & 33.6 & 9.38 & 40.68 & 0.38 & $ 7.5\times10^{-3}$\\
NGC 2992 & 146.4251875 & $-$14.32636110   & Sy2   & 0.0077 & 33.2 & 8.33 & 40.11 & 0.10 & $ 2.0\times10^{-3}$\\
NGC 3079 & 150.4911624 & $+$55.67984716   & Sy2   & 0.0037 & 15.9 & 6.38 & 40.04 & 0.09 & $ 1.7\times10^{-3}$\\
NGC 3516  & 166.6977634 & $+$72.56869399      &  Sy1.2 & 0.0088 & 37.9 & 7.39 & 39.63    &  & \\
NGC 4151 & 182.6357333 & $+$39.40585098   & Sy1   & 0.0033 & 14.2 & 7.55 & 41.21 & 1.27 & $ 2.5\times10^{-2}$\\
NGC 4235 & 184.2911746 & $+$7.19157594  & Sy1   & 0.0080 & 34.5 & 7.55 & 40.94 & 0.68 & 1.4$\times10^{-2}$\\
NGC 4388 & 186.4450833 & $+$12.66206940   & Sy2   & 0.0039 & 16.8 & 6.94 & 41.57 & 2.93 & 5.9$\times10^{-2}$\\
NGC 4593 & 189.9143480 & $-$5.34417639   & Sy1   & 0.0090 & 38.8 & 6.88 & 40.65 & 0.35 & 7.0$\times10^{-3}$\\
NGC 5290 & 206.3298250 & $+$41.71235376   & Sy2   & 0.0086 & 37.1 & 7.78 & 40.12 & 0.10 & 2.1$\times10^{-3}$\\
NGC 5506 & 213.3119838 & $-$3.20765470   & Sy1.9 & 0.0062 & 26.7 & 6.96 & 41.18 & 1.19 & 2.4$\times10^{-2}$\\
\enddata
\label{tab:sample}
\tablecomments{ICRS (a catalogue of extragalactic radio sources) coordinates were retrieved from the astropy.coordinates package\footnote{\url{https://docs.astropy.org/en/stable/api/astropy.coordinates.get_icrs_coordinates.html}}. $M_\mathrm{BH}$ values were estimated from the M$-\sigma^*$ correlation when other more direct measurements were not available and the distances were obtained from redshift measurements (as explained in Paper I).}
\end{deluxetable*}

\section{Methodology} \label{sec:Methodology}
\subsection{Sample Selection} \label{subsection: Sample Selection}

\begin{figure}
\includegraphics[width=\columnwidth]{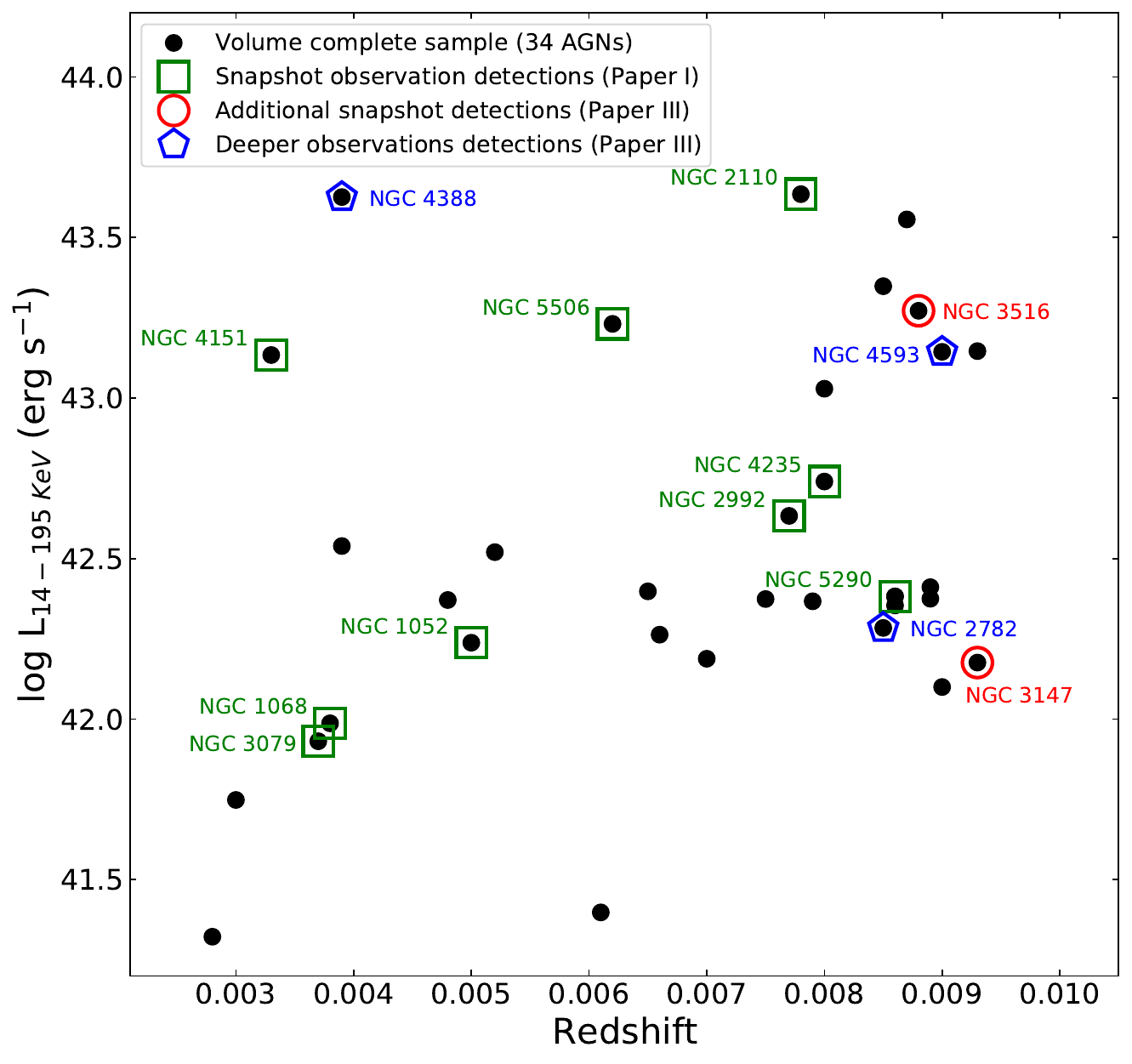}
\caption{Hard X-ray (14$-$195 keV) luminosity vs. redshift for our volume complete sample of 34 nearby AGNs from the 105-month \emph{Swift} BAT catalog. The green squares are the sources detected in our initial snapshot campaign (Paper~I), whereas the red circles are the detected sources from our additional snapshot observation (Paper~III). The blue pentagons are a subset of the Paper I sample detected in our follow-up deep integrations observation discussed in Paper~III. However, NGC 2782 and NGC 3147 were dropped from our final list of sample sources (see Section~\ref{sub:obs}).} 
\label{fig:sample}
\end{figure}

In Paper I, we had the opportunity to observe 25 AGNs with the VLBA at milliarcsecond (sub-parsec) scales in C band/6 cm and detected only 9 sources. Later, a follow-up observing program recovered 5 more sources (as described in Paper III). Three of these were detected in a deeper 4-hour on-source integration time survey, which followed up on previous non-detections. The remaining two sources were detected during observations that completed the volume-limited sample of 34 AGNs. Here, we selected 12 sources that were detected in our previous analyses, which correspond to all detected sources except for NGC 2782 and NGC 3147, as detailed in Section~\ref{subsection: VLBA}, for a multi-wavelength study at L, C, X, and K bands with the VLBA. We illustrate our selection method in Figure~\ref{fig:sample}, and sample sources, along with their global properties, are listed in Table \ref{tab:sample}.

\begin{deluxetable*}{lcchcccc} 
\label{tab:Observation}
\tablecaption{Details of VLBA Multiband Observation Segments}
\tablehead{\colhead{Segment} & \colhead{Target} & \colhead{Obs. Band} & 
& \colhead{Source Scan Time} & \colhead{Phase Calibrator}& \colhead{Separation}&\colhead{Obs. Date Range (2021)}\\%
[-0.2cm]
\colhead{}& \colhead{}       &  \colhead{}	& 
& \colhead{(min)}  & \colhead{}& \colhead{(deg)}&  \colhead{}}
\startdata
UC002A & NGC 3079 & L, C, X &1.4, 4.1, 8.4 & 55, 27, 63 & J1014+5503& 1.85 & May-03 23:24 $-$ May-04 05:27\\
\hline
UC002C & NGC 4388 & L, C, X &1.4, 4.1, 8.4 & 55, 27, 62 & J1225+1253& 0.28  & Apr-14 03:27 $-$ Apr-14 09:22\\
\hline
UC002D & NGC 4151 & K &21.7& 230 & J1206+3941&0.81 & May-20 23:47 $-$ May-21 07:35\\
UC002E & NGC 4151 & L, C, X &1.4, 4.1, 8.4 & 54, 27, 62  & J1206+3941&0.81 & May-21 23:43 $-$ May-22 05:45\\
\hline
UC002F & NGC 5290 & K &21.7& 230 & J1359+4011& 3.10 & Jun-18 23:07 $-$ Jun-19 07:30\\
UC002G & NGC 5290 & L, C, X &1.4, 4.1, 8.4 &55, 27, 62 & J1359+4011& 3.10 & Jun-18 00:11 $-$ Jun-18 06:35\\
\hline
UC002I & NGC 5506 & L, C, X &1.4, 4.1, 8.4 & 54, 27, 62 & J1404$-$0130&2.72 & Jun-20 00:33 $-$ Jun-20 06:37\\
\hline
UC002L & NGC 1052 & K &21.7& 245 & \nodata&\nodata & Jul-30 11:09 $-$ Jul-30 11:09\\
UC002M & NGC 1052 & L, C, X &1.4, 4.1, 8.4 & 56, 28, 64 & \nodata&\nodata & Jul-29 11:13 $-$ Jul-29 14:33\\
\hline
UC002N & NGC 3516 & C, K &4.1, 21.7& 27, 230 & J1048+7143&1.64 & Jul-30 16:38 $-$ Jul-31 02:44\\
\hline
UC002O & NGC 1068 &  K &21.7 & 230 & 
J0239$-$0234&2.72 & Aug-05 08:47 $-$ Aug-05 16:40\\
UC002P & NGC 1068 & L, C, X &1.4, 4.1, 8.4& 55, 27, 62 & 
J0239$-$0234&2.72  & Aug-06 09:41 $-$ Aug-06 15:45\\
\hline
UC002Q & NGC 2110 & K &21.7 & 230 & J0553$-$0840&1.27 & Aug-28 10:26 $-$ Aug-28 18:10\\
UC002R & NGC 2110 & L, C, X &1.4, 4.1, 8.4& 54, 27, 62 & J0553$-$0840&1.27 & Aug-29 11:36 $-$ Aug-29 17:35\\
\hline
UC002T & NGC 2992 & K &21.7& 230 & J0941$-$1335&1.34 & Oct-17 10:57 $-$ Oct-17 18:55\\
UC002U & NGC 2992 & L, C, X &1.4, 4.1, 8.4& 54, 27, 62 & J0941$-$1335&1.34 &Oct-19 11:49 $-$ Oct-19 17:48\\
\hline
UC002W & NGC 4235 & L, C, X &1.4, 4.1, 8.4 & 54, 27, 62 & J1214+0829&1.40 & Dec-23 10:28 $-$ Dec-23 16:28\\
UC002X & NGC 4235 & K &21.7& 230 & J1214+0829&1.40 & Dec-17 09:32 $-$ Dec-17 17:30\\
\hline
\hline
&&&&&&& Obs. Date Range (2022)\\
\hline
UC002Y & NGC 4593 & K &21.7 & 230 & J1248$-$0632&2.48 & Jan-06 08:38 $-$ Jan-06 16:37\\
UC002Z & NGC 4593 & L, C, X &1.4, 4.1, 8.4& 54, 27, 62 &J1248$-$0632&2.48 & Jan-04 09:46 $-$ Jan-04 15:49\\
\hline
UC002AA$^{*}$ & NGC 3079 & K &21.7& 288 & J0957+5522&0.68 & Aug-31 13:33 $-$ Aug-31 23:23\\
\hline
UC002AB$^{*}$ & NGC 4388 & K &21.7& 230 & J1225+1253&0.28 & Oct-18 13:29 $-$ Oct-18 21:12\\
\hline
UC002AC$^{*}$ & NGC 5506 & K &21.7& 230 & J1411$-$0300&0.58 & Oct-03 16:23 $-$ Oct-04 00:12\\
\hline
UC002AD$^{*}$ & NGC 3516 & L, X & 1.4, 8.4& 55, 62 & J1048+7143&1.64  & Aug-04 20:48 $-$ Aug-05 00:51\\
\enddata
\tablecomments{The rest frequencies for L, C, X, and K bands are 1.376, 4.112, 8.376, and 21.712 GHz, respectively. The phase calibrator sources were taken from the ICRF3 S/X catalog~\citep{Charlot_2020...644A.159C} and the RFC\footnote{\url{http://astrogeo.smce.nasa.gov/rfc/}} (only for J1014+5503 and J1411-300).\\
$^{*}$VLBA follow-up observation segments.}
\end{deluxetable*} 

\subsection{VLBA Data}  \label{subsection: VLBA}
\subsubsection{Observations}\label{sub:obs}
Using the radio interferometry sensitivity equation and verifying with the European VLBI Network’s (EVN\footnote{\url{http://old.evlbi.org/cgi-bin/EVNcalc.pl}}) online calculator, on-source integration times necessary to produce images with thermal noise levels of $\sim$25 $\mu$Jy/beam with dual polarization are estimated as 53 minutes at L-band, 26 min at C-band, 63 min at X-band, and 4 hours at K-band. We employed phase referencing for all targets except for NGC 1052, which showed an S/N ratio of $\sim$ 200 at 6 GHz in Paper I. 
We adopt the nominal absolute flux density calibration error of 10$\%$ and do not require separate flux density calibration scans. 
We designed an efficient observing program optimized for the science goals while allowing flexibility for easy incorporation into the dynamic scheduling process. We split the K-band observations for each target into two 4-hour segments, which can be scheduled dynamically as needed due to the requirement for good weather. For all remaining L-band to X-band observations, we combined them into a third key file, switching between short sets of each receiver setting spread evenly throughout a single track to maximize UV coverage. We set up the dynamic scheduling constraints such that all three schedule files per source were completed within $\sim$24-48 hours to ensure quasi-simultaneous coverage across all the radio bands. NGC 2782, the weakest detection from our deeper observations (Paper III), was dropped because it was five times fainter than other detections in C-band and only four times our estimated theoretical best RMS noise floor. In addition, NGC 2782 was dropped because of being too close to the Sun during the observation period. On the other hand, NGC 3147 was observed without any phase calibrator, similar to NGC 1052, because it showed an S/N ratio of $\sim$ 150 at 6 GHz in Paper III and was expected to be bright enough to fringe fit on itself, but it failed the fringe fitting task and thus, was dropped from our final list of the sample sources. The details of VLBA multiband observation segments, including scan time, date of observation, and list of phase calibrators, are shown in table~\ref{tab:Observation}.

\subsubsection{Calibration and Imaging}
We used National Radio Astronomy Observatory (NRAO) Astronomical Image Processing System \citep[\textsc{aips};][]{Van_1996ASPC..101...37V} release 31DEC19 to calibrate our VLBA data. Each frequency dataset was calibrated independently by a target -- phase calibrator pair (except for NGC 1052).
Bad data were flagged, and calibration was performed using the standard \textsc{aips} VLBA procedures, following the prescription outlined in Paper~I. 
We used the \textsc{aips} task {\sc imagr} to make images of the calibrators and sources, cleaning the images until the rms approached three times the theoretical thermal noise limit. We used successive iterations of phase-only self-calibration by using \textsc{aips} task {\sc calib}. The process used models produced in a previous image applied to the next iteration to improve the source phase. Self-calibration was applied to brighter sources, including NGC 1052, NGC 3079, and NGC 5506 (see Table~\ref{tab:image_properties} for details). In cases where self-calibration was not feasible for some or all of the four band images, a correction factor was applied to the peak intensities to prevent coherence loss~\citep{Mart_2010A&A...515A..53M}. A correction factor was obtained for each band by comparing other frequency datasets of the same source with and without self-calibration. The resulting value ($\sim$10-30$\%$ of peak intensities based on the frequency band) was then added to the total uncertainty calculation for observations without self-calibration. Later we applied amplitude and phase self-calibration to improve amplitude when necessary. Self-calibration was iterated until the image signal-to-noise (S/N) ratio improved and the ratio between bad and good solutions in the {\sc calib} task was~$\leq$10\%.
We achieved an S/N ratio of $>$10 for most of our detections in all bands except for a few sources where the S/N ratio was limited to $\leq$5 at higher frequencies, especially in the K band.  For more information on our calibration and imaging procedures, please see Sections 2.2.1 and 2.2.2, respectively, in Paper~I.

\startlongtable
\begin{deluxetable*}{lccccHcrc} 
\label{tab:image_properties}
\setlength{\tabcolsep}{4pt}
\tablecaption{Radio Image Properties}
\tablehead{\colhead{Target} & \colhead{Frequency} & \colhead{Restoring Beam}                 & \colhead{Beam Angle} & \colhead{RMS} & 
&\colhead{F$^\mathrm{peak}_\mathrm{Gauss}$}& \colhead{S$^\mathrm{int}_\mathrm{Gauss}$} & Self-calibration\\%
[-0.2cm]
& \colhead{Band}       &  \colhead{($\alpha \times \delta$; mas)}	& \colhead{(deg)} & \colhead{(mJy bm$^{-1}$)}  & 
& \colhead{(mJy bm$^{-1}$)}&   \colhead{(mJy)} & Applied}
\startdata
NGC 1052 & L & 16.32$\times$5.53 &$-$3.05 & 0.82 & 279.0 & $273.5\pm1.2$ & $461\pm8$ & yes\\
 & C & 8.53$\times$4.25 &$-$3.25 & 6.50 & 916.9 & $912\pm15$ & $1131\pm30$ & yes\\
 & X & 4.76$\times$3.42 & 10.81 & 8.00 & 1196.3  & $1162\pm47$&$1610\pm100$& yes\\
 & K & 1.62$\times$0.38 &$-$20.15& 3.50 &154.3 & $137.2\pm6.9$&$636\pm44$& yes\\
 \hline
NGC 1068 & L & 19.04$\times$9.12 &$-$0.73 & 0.11 & $<$0.55&$<$0.55&\nodata& \nodata\\
 & C &7.64$\times$4.27 &$-$5.53 & 0.05 & 0.71 &$0.72\pm0.06$&$2.60\pm0.28$& no\\
 & X & 4.18$\times$3.16 &$-$0.60 & 0.04 & 0.36&$0.37\pm0.02$&$0.39\pm0.04$& no\\
 & K & 1.02$\times$0.32 &$-$13.64 & 0.04 & $<$0.20 &$<$0.20&\nodata& \nodata\\
 \hline
NGC 2110 & L & 13.63$\times$4.84&$-$9.79 & 0.16 & 4.60  &$4.53\pm 0.15$& $12.17\pm0.62$ & no\\
 & C & 8.25$\times$4.40 & 2.63 & 0.12 & 39.72  &$40.81\pm 0.37$&$42.76\pm 0.70$& yes\\
 & X & 5.01$\times$3.34 &3.05 & 0.06 & 29.34  &$29.07\pm 0.17$&$35.48\pm 0.34$& yes\\
 & K & 1.33$\times$0.34&$-$14.44 & 0.05 & $<$0.25  &$<$0.25&\nodata& \nodata\\
 \hline
NGC 2992 & L & 14.84$\times$4.96 &$-$7.61 & 0.09 & 2.22 &$2.04\pm0.13$&$3.81\pm0.40$ & no\\
 & C  & 8.63$\times$4.10 &3.22 & 0.10 & 1.72  &$1.78\pm0.07$&$1.83\pm0.13$& no\\
 & X & 5.18$\times$3.01 &3.72 & 0.10 & 1.52 &$1.56\pm0.10$&$1.37\pm0.17$& no\\
 & K & 1.69$\times$0.42 &$-$16.29 &0.09 & 0.59  &$0.62\pm0.05$ & $0.74\pm0.15$& no\\
 \hline
NGC 3079 & L & 9.81$\times$4.95 &$-$11 & 0.25 & 5.87 (A)&$6.33\pm0.27$ (A)& $9.5\pm0.7$(A)& yes\\
 &  &  & & & $<$1.25 (B) & $<$1.25 (B)&\nodata (B)& yes\\
 & C & 5.13$\times$4.15 &8.53 & 0.13 & 51.80 (A)  &$53.10 \pm 0.85$ (A)&$85.1 \pm 2.0$ (A)& yes\\
 &  &  & & & 11.24 (B) & $11.31\pm0.10$ (B)&$13.27\pm0.19$ (B)& yes\\
 & X & 3.59$\times$3.12 &17.12 & 0.16 & 22.40 (A) & $22.92\pm0.75$ (A)& $47.9\pm2.2$ (A)& yes\\
 &  &  & & & 28.91 (B) & $28.32\pm0.56$ (B)&$35.2\pm1.1$ (B)& yes\\
 & K & 0.55$\times$0.39 &$-$18.8 & 0.06 & $<$0.30 (A)& $<$0.30 (A)& \nodata(A)& \nodata\\
 &  &  & & & $<$0.30 (B) & $<$0.30 (B)&\nodata(B)& \nodata\\
 \hline
NGC 3516 & L & 10.28$\times$4.71 &$-$17.07 & 0.11 & 0.48 &$2.09\pm0.16$&$3.26\pm0.47$& yes\\
 & C & 5.29$\times$4.77 &$-$3.22 & 0.04 & 0.54  &$0.52\pm0.03$&$0.72\pm0.07$& no\\
 & X   & 3.52$\times$2.66 &$-$5.31 & 0.05 & 0.53  &$0.49\pm0.02$&$0.82\pm0.06$& no\\
 & K & 2.11$\times$1.81 &25.33 & 0.09 & $<$0.45 &$<$0.45&\nodata& \nodata\\
 \hline
 NGC 4151 & L & 11.64$\times$5.64 &$-$7.08 & 0.09 & 9.02 &$8.02\pm0.53$&$24.7\pm2.0$& yes\\
 & C & 5.53$\times$4.07 &$-$13.81 & 0.07 & 9.69 & $8.77\pm0.56$&$28.3\pm2.3$& yes\\
 & X & 3.72$\times$3.50 &37.38 & 0.12 & 5.48 & $4.84\pm0.29$&$21.7\pm1.6$& yes\\
 & K & 1.97$\times$0.86 &43.37 & 0.06 & 0.26  & $0.27\pm0.03$&$0.79\pm0.18$& no\\
 \hline
NGC 4235 & L & 12.99$\times$4.98 &$-$10.68 & 0.06 & 1.46&$1.43\pm0.07$ &$1.81\pm0.16$& no\\
 & C & 9.34$\times$5.24 &42.02 & 0.13 &3.08 & $3.06\pm0.15$& $3.21\pm0.28$& no\\
 & X & 4.83$\times$3.49 &41.99& 0.11 & 2.11 & $1.97\pm0.10$& $3.59\pm0.28$& no\\
& K & 2.37$\times$1.35 &21.18& 0.09 & $<$0.45&$<$0.45&\nodata & \nodata\\ 
\hline
NGC 4388 & L & 14.40$\times$4.39 &$-$15.57 & 0.03 & 0.19 & $0.19\pm0.02$ &$0.22\pm0.05$& no\\
 & C & 8.65$\times$4.19 &$-$19.22 & 0.04 & 0.24  &$0.23\pm0.02$&$0.59\pm0.08$& no\\
 & X & 3.94$\times$3.28 &1.81 & 0.05 &0.26 & $0.25\pm0.02$&$0.41\pm0.06$& no\\
& K & 1.27$\times$0.53 &$-$11.14 & 0.08 &$<$0.40 &$<$0.40&\nodata & \nodata\\
 \hline
NGC 4593 & L & 15.30$\times$5.17 &$-$7.01 & 0.06 & $<$0.30 & $<$0.30&\nodata & \nodata\\
 & C & 7.90$\times$4.36 &27.39 & 0.06 & 0.57 & $0.65\pm0.04$&$0.77\pm0.08$& no\\
 & X & 5.58$\times$3.88 &31.37 & 0.05 & 0.50  & $0.47\pm0.05$&$0.88\pm0.14$& no\\
 & K & 2.30$\times$0.53 &$-$14.13 &0.04 & 0.34  & $0.29\pm0.03$&$0.64\pm0.11$& no\\
 \hline
 NGC 5290 & L & 9.73$\times$4.60 &9.97 & 0.03 & 0.32  &$0.29\pm0.05$&$0.66\pm0.14$& no\\
 & C & 4.82$\times$3.64 &$-$3.97 & 0.03 & 4.38 &$4.75\pm0.02$&$4.83\pm0.04$& yes\\
& X & 3.88$\times$3.34 &19.61 & 0.06 & 6.26& $6.33\pm0.07$&$6.74\pm0.13$& yes\\
 & K & 0.71$\times$0.38 &$-$7.60 & 0.04 & 0.51  &$0.49\pm0.05$&$1.62\pm0.33$& no\\
 \hline
NGC 5506 & L & 16.12$\times$5.36 &$-$11.79 & 0.35 & 19.17 (B0) & $24.3\pm1.5$ (B0)& $33.5\pm3.7$ (B0)& yes\\
 & &  & &  & 5.99 (B1) & $8.95\pm0.38$ (B1)& $31.1\pm3.2$ (B1)& yes\\
 & C & 7.70$\times$4.36 &6.74 & 0.21 & 43.67 (B0) &$44.27\pm0.39$ (B0)&$48.83\pm0.76$ (B0)& yes\\
 & &  & &  & 6.85 (B1) & $6.75\pm0.34$ (B1)&$28.6\pm1.8$ (B1)& yes\\
 & X & 4.60$\times$3.23 &0.15 & 0.12 & 27.66 (B0) &$27.43\pm0.26$ (B0)&$34.79\pm0.54$ (B0)& yes\\
 & &  & &  & 1.20 (B1) & $1.13\pm0.07$ (B1)&$9.13\pm0.63$ (B1)& yes\\
 & K & 1.51$\times$0.26 &$-$22.52 & 0.21 & &$<$1.05 (B0) & \nodata (B0) & \nodata\\
 &  &  & &  & & $<$1.05 (B1) &\nodata (B1) & \nodata\\
\enddata
\tablecomments{Peak intensities (F$^\mathrm{peak}_\mathrm{Gauss}$) and integrated flux densities (S$^\mathrm{int}_\mathrm{Gauss}$) are measured from CASA's 2-D Gaussian model fitting algorithm. 
For nondetections, upper limits were measured using a 5$\sigma$ intensity limit above the rms. Beam angle is the beam's major axis position angle measured from north towards the east.}
\end{deluxetable*}

\section{Results} \label{sec:res}

\begin{figure*}[htp]
\centering
\vspace{1cm}
\large \textbf{NGC 1052}\par\medskip
\includegraphics[clip, trim=5cm 15.5cm 5cm 4.3cm, width=0.67\textwidth]{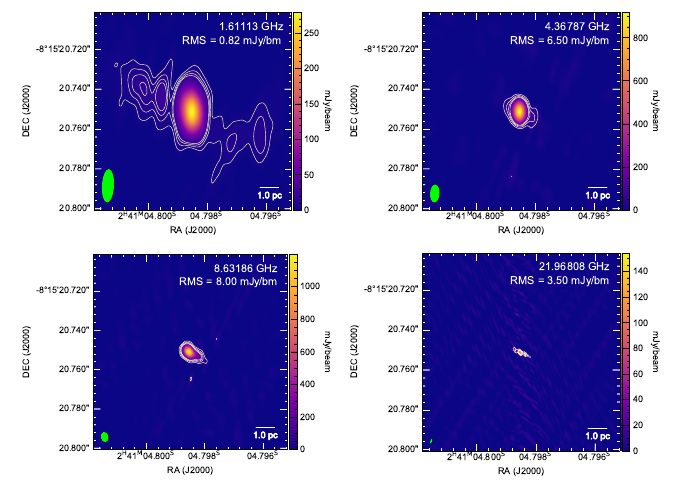}
\includegraphics[clip,trim=0.2cm 0cm 0cm 0.3cm, width=0.3\textwidth]{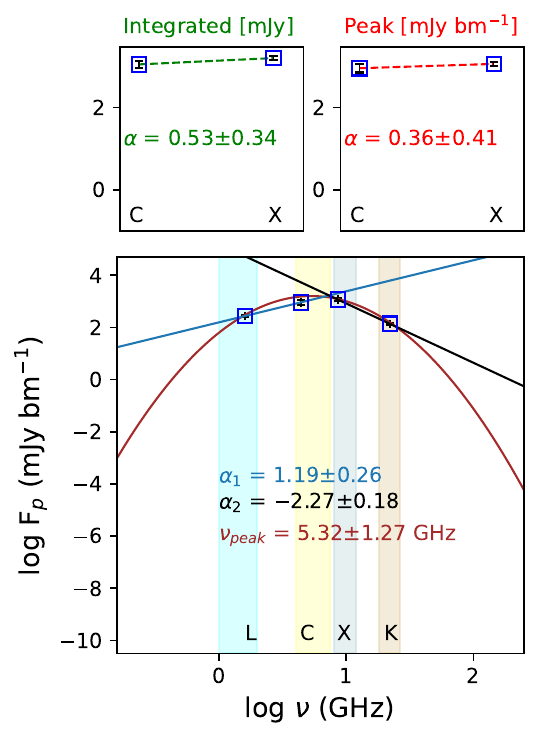}\\
\vspace{1cm}
\large \textbf{NGC 1068}\par\medskip
\includegraphics[clip, trim=5cm 15.3cm 5cm 4.3cm, width=0.67\textwidth]{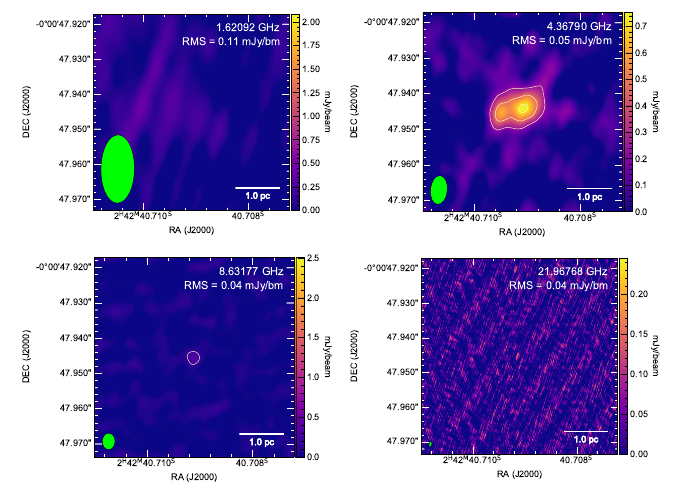}
\includegraphics[clip,trim=0.2cm 0cm 0cm 0.3cm, width=0.3\textwidth]{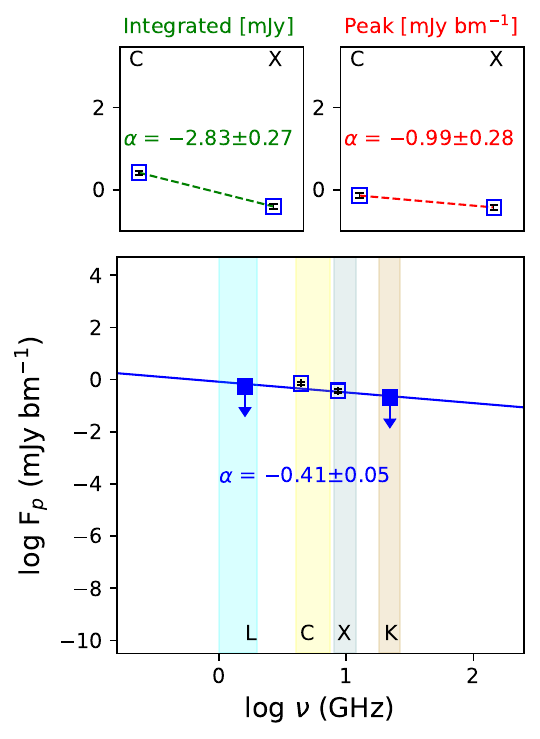}\\
\caption{Multiband (1.6 to 22 GHz) parsec-scale VLBA radio observations. The outermost contour represents the 5$\sigma$ flux limit above RMS, and the interior contours increase as $\sigma \times$(10, 15, 20). The green ellipses to the lower left of each frame represent the synthesized beam size for that observation. The right panel represents the Spectral Energy Distribution (SED) for each source using the peak intensities of each frequency image. The nondetection upper limits (5$\sigma$) are denoted by a downward arrow. The two plots in the upper panel represent the spectral shape only between 4.4 GHz (C band) and 8.6 GHz (X band) after applying \textit{UVTAPER} on both images. }
\label{fig:multiband_image}
\end{figure*}

Table~\ref{tab:image_properties} lists the multi-wavelength radio observation properties for our final calibrated images displayed in Figure \ref{fig:multiband_image} for all sources. Along with four frequency band images, we added a plot of spectral energy distribution (SED) using peak intensities (log F$^\mathrm{peak}_\mathrm{Gauss}$) and observed frequencies (log $\nu$) in Figure \ref{fig:multiband_image}. 
Peak intensity values for all four bands were measured from CASA’s 2-D Gaussian model fitting algorithm. We used simple functions to characterize the SEDs
in the log-log plane of peak intensity and frequency: (1) standard power law and (2) best-fitting quadratic/parabola or curved power-law model (a Gaussian in non-logarithmic space). In addition, we applied a piecewise linear regression model for sources fitted with a curved power-law model to demonstrate the spectral shape in lower and higher frequency ends separately. For a single power law spectrum,  flat- and steep-spectrum
radio sources are defined to have spectral indices smaller or larger than a
particular limiting value, typically $\alpha_{lim} \sim 0.5$~\citep{Tadhunter_2016A&ARv..24...10T}. Following the spectral index sign convention used in this paper ($S_\mathrm{\nu}\propto \nu^\mathrm{+\alpha}$), we define a steep spectrum as having $\alpha$ $\leq-$0.5, a flat one as having $-0.5 < \alpha \leq 0$, and an inverted spectrum as having $\alpha >$ 0. To determine whether the radio-emitting region is an extended or a point-like source, we have added two additional SED plots for each source using peak and integrated flux density measurements only from C and X bands. The difference between integrated flux value and peak intensity can provide information about the extension of the radio-emitting region. However, before making such comparisons, we applied {\sc uvtaper} in \textsc{aips} task {\sc imagr} to specify widths ($\sim$ 50000 k$\lambda$) in U and V directions of the Gaussian function, which made their beam sizes comparable; thus, their integrated flux densities can provide a meaningful comparison in SED fitting. In the following, we summarize the specific details of source images and SED plots. We have also added the coordinates obtained from our phase-referenced VLBA measurements, with precision truncated to the formal uncertainties in the phase reference source catalogs (please see Table~\ref{tab:Observation}).\\

\noindent{\textbf{NGC 1052 [2$^{\rm h}$41$^{\rm m}$04$\fs$7985, $-$08$\degr$15$\arcmin$20$\farcs$751]}}: All four band images revealed a bright center with extended east-west double-sided collimated outflows. Similar double-sided outflows were observed in multiple previous VLBA studies of this source~\citep{Kameno_2003PASA...20..134K, Sawada_2008ApJ...680..191S,Baczko_2022A&A...658A.119B}. NGC 1052 is the only source in our sample showing a prominent double-sided jet-like collimated structure extending outward from the central region, even with the high-resolution ($\leq$1 pc) K-band observation. The four-frequency SED plot shows a GHz-peaked (curved) spectrum with a peak near $\sim$ 5 GHz. The piecewise regression model yields two different spectral slopes ($\alpha$) with two linear functions connected to a ``knot value'' indicating steep spectral indices at high frequencies and an inverted spectral slope at lower frequencies. The two additional SED sub-plots using only C and X bands show similar spectral shapes, which indicates that the bright central region has a compact point-like structure.\\  

\noindent{\textbf{NGC 1068 [2$^{\rm h}$42$^{\rm m}$40$\fs$70890, 00$\degr$00$\arcmin$48$\farcs$0062]}}: This source is detected in C and X bands but not in L and K. Note that NGC 1068 suffers from severe short spacing issues due to the bright, extended radio continuum emission and that the non-detection at L band may be due to these effects. However, all four band peak intensities, including 5$\sigma$ nondetections upper limits, show a flat (typically, if $\alpha$ $> -$0.5 ) spectrum.  A little extended structure was detected in the C band image; as a result, a steeper spectrum was seen when $\alpha$ was measured only between C and X bands integrated flux densities. \\ 

\noindent{\textbf{NGC 2110 [5$^{\rm h}$52$^{\rm m}$11$\fs$3763, $-$07$\degr$27$\arcmin$22$\farcs$519]}}: L, C, and X bands images revealed a bright point source at the center. No emission was detected in the K band image, so we used the nondetection upper limit for the SED. A GHz-peaked spectrum is found for NGC 2110 with a steeper spectrum in a higher frequency regime beyond $\sim$ 5 GHz. The C$-$X band SED sub-plots are similar with much flatter indices ($\alpha$ $> -$0.5) for peak intensities and integrated flux densities, suggesting a central point source. \\

\noindent{\textbf{NGC 2992 [9$^{\rm h}$45$^{\rm m}$41$\fs$94437, $-$14$\degr$19$\arcmin$34$\farcs$599]}}: NGC 2992 is detected in all four bands and shows a bright, compact radio emission at the center. The point-like nature of this source can also be deduced from the C to X band flat SEDs using both peak intensities and integrated flux densities. In addition, the four-frequency SED plot shows a flat spectrum with a spectral index, $\alpha > -$0.5. \\

\renewcommand{\thefigure}{\arabic{figure} (Cont.)}
\addtocounter{figure}{-1}

\begin{figure*}[htp]
\centering
\vspace{1cm}
\large \textbf{NGC 2110}\par\medskip
\includegraphics[clip, trim=5cm 15.3cm 5cm 4.3cm, width=0.67\textwidth]{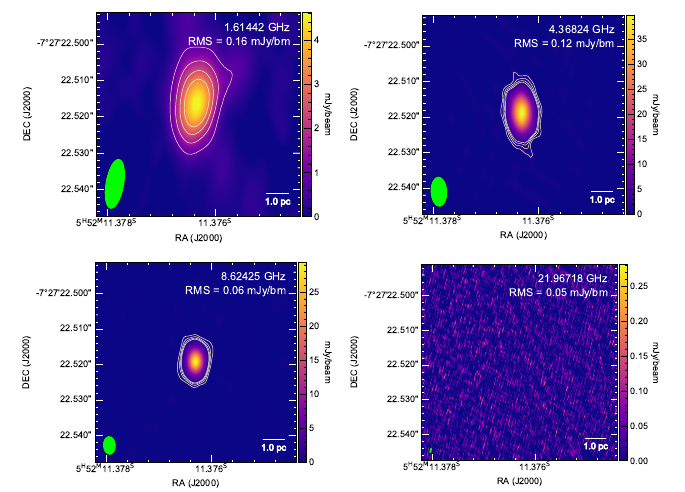}
\includegraphics[clip,trim=0.2cm 0cm 0cm 0.3cm, width=0.3\textwidth]{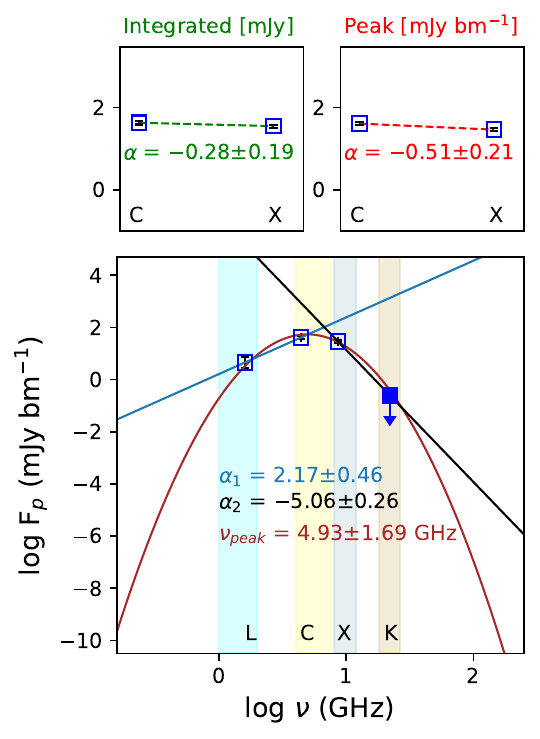}\\
\vspace{1cm}
\large \textbf{NGC 2992}\par\medskip
\includegraphics[clip, trim=5cm 15.3cm 5cm 4.3cm, width=0.67\textwidth]{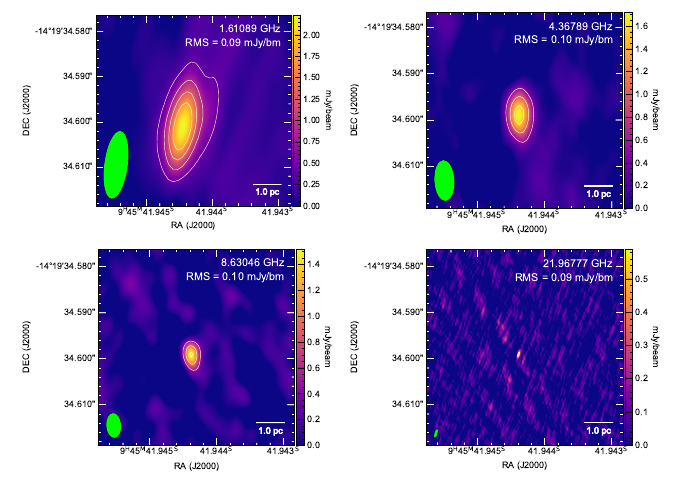}
\includegraphics[clip,trim=0.2cm 0cm 0cm 0.3cm, width=0.3\textwidth]{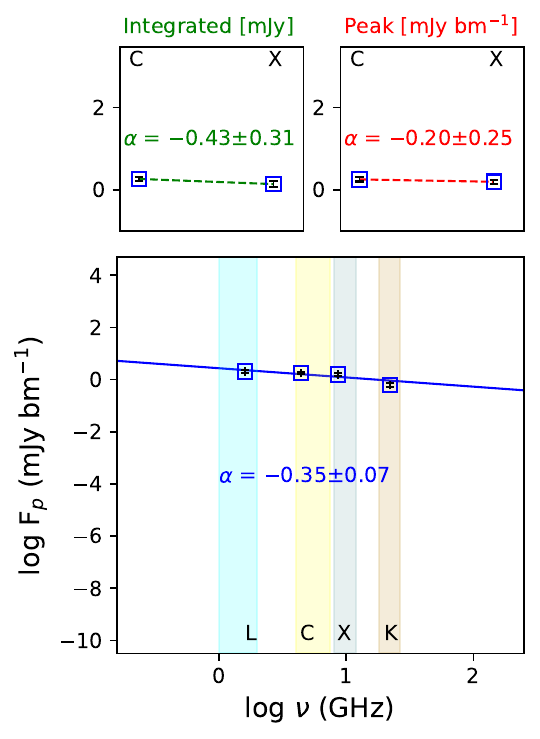}\\
\caption{}
\end{figure*}

\addtocounter{figure}{-1}

\begin{figure*}[htp]
\centering
\vspace{1cm}
\large \textbf{NGC 3079}\par\medskip
\includegraphics[clip, trim=5cm 15.3cm 5cm 4.3cm, width=0.67\textwidth]{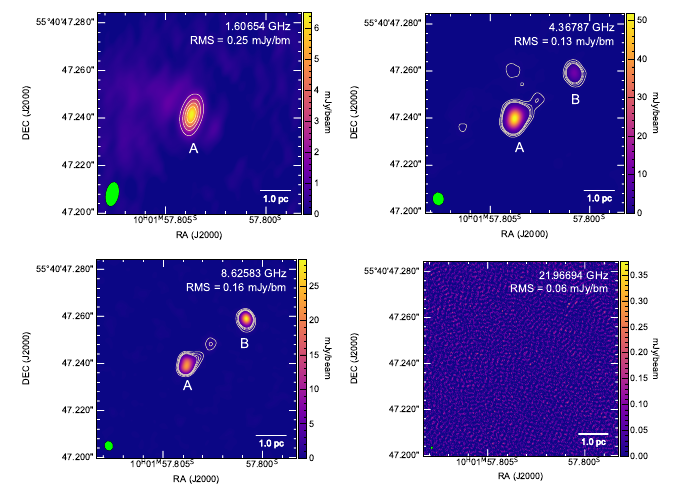}
\includegraphics[clip,trim=0.2cm 0cm 0cm 0.3cm, width=0.3\textwidth]{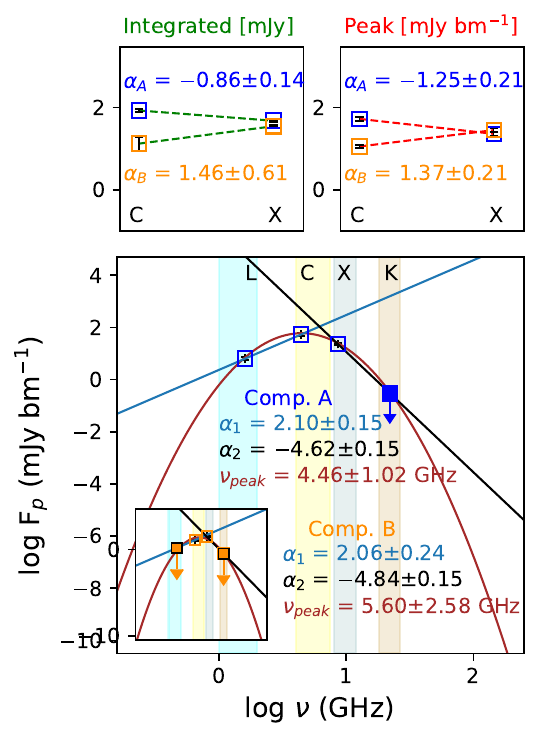}\\
\vspace{1cm}
\large \textbf{NGC 3516}\par\medskip
\includegraphics[clip, trim=5cm 15.3cm 5cm 4.3cm, width=0.67\textwidth]{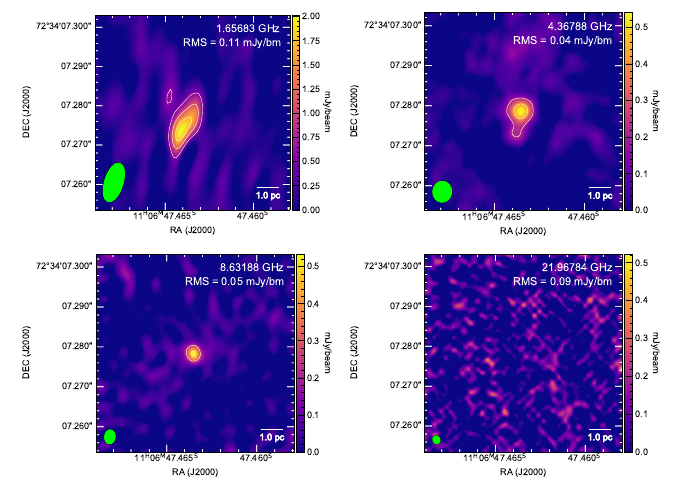}
\includegraphics[clip,trim=0.2cm 0cm 0cm 0.3cm, width=0.3\textwidth]{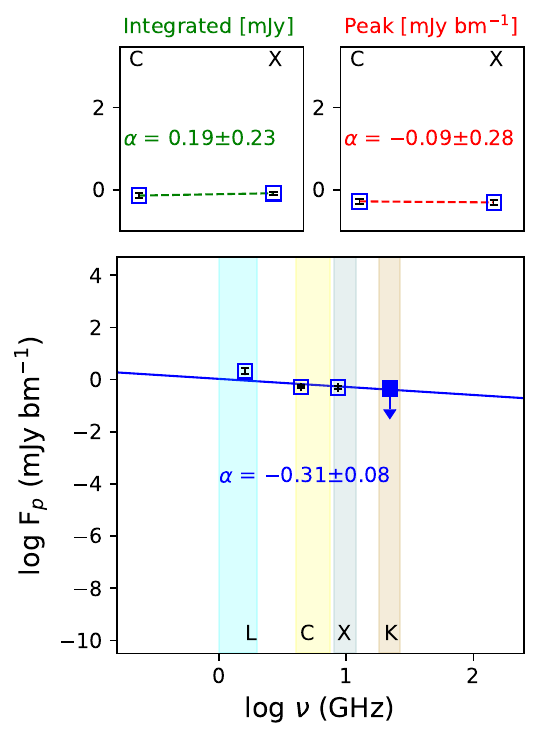}\\
\caption{}
\end{figure*}

\addtocounter{figure}{-1}

\begin{figure*}[htp]
\centering
\vspace{1cm}
\large \textbf{NGC 4151}\par\medskip
\includegraphics[clip, trim=5cm 15.3cm 5cm 4.3cm, width=0.67\textwidth]{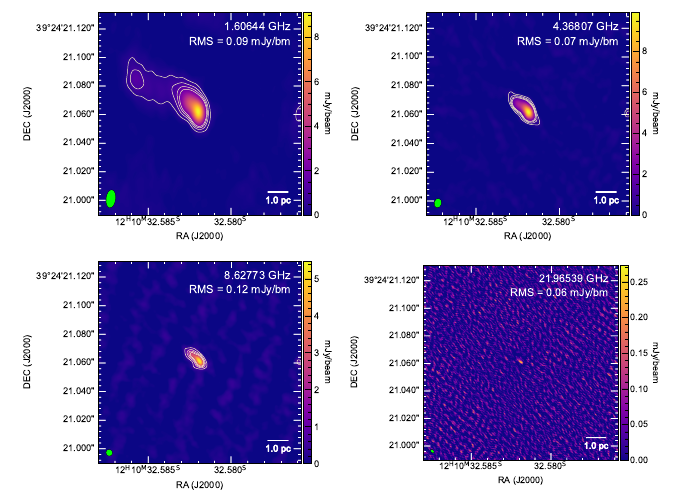}
\includegraphics[clip,trim=0.2cm 0cm 0cm 0.3cm, width=0.3\textwidth]{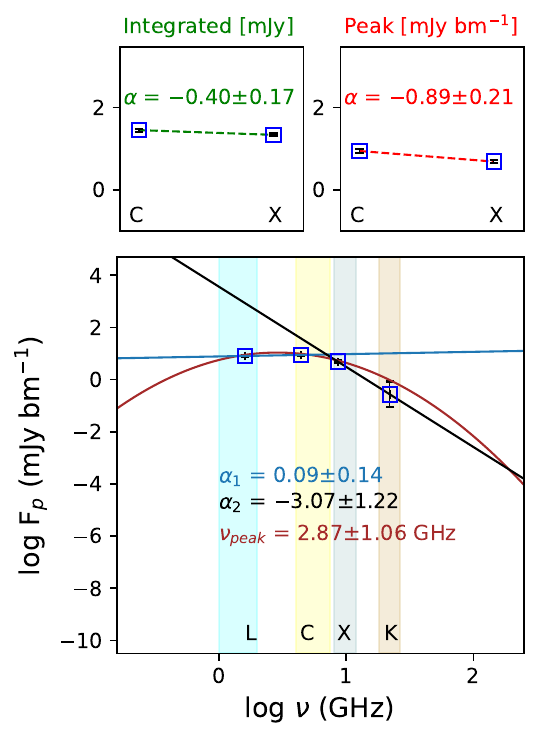}\\
\vspace{1cm}
\large \textbf{NGC 4235}\par\medskip
\includegraphics[clip, trim=5cm 15.3cm 5cm 4.3cm, width=0.67\textwidth]{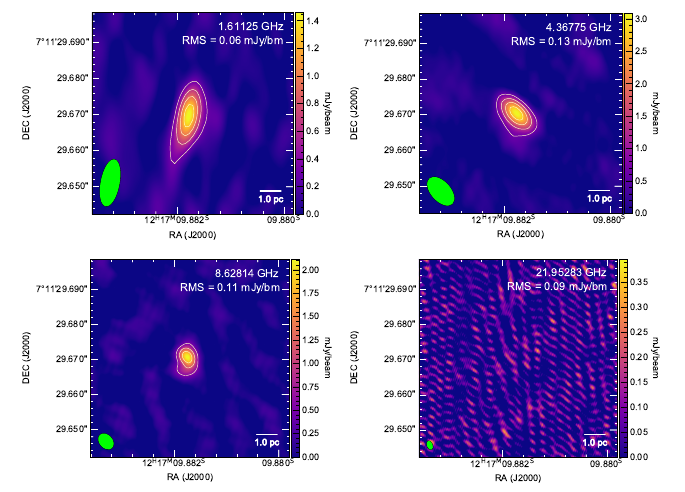}
\includegraphics[clip,trim=0.2cm 0cm 0cm 0.3cm, width=0.3\textwidth]{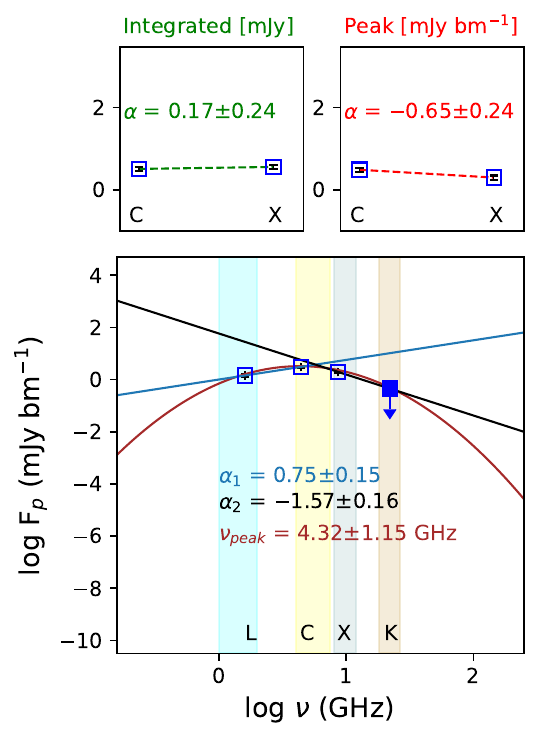}\\
\caption{}
\end{figure*}

\addtocounter{figure}{-1}

\begin{figure*}[htp]
\centering
\vspace{1cm}
\large \textbf{NGC 4388}\par\medskip
\includegraphics[clip, trim=5cm 15.3cm 5cm 4.3cm, width=0.67\textwidth]{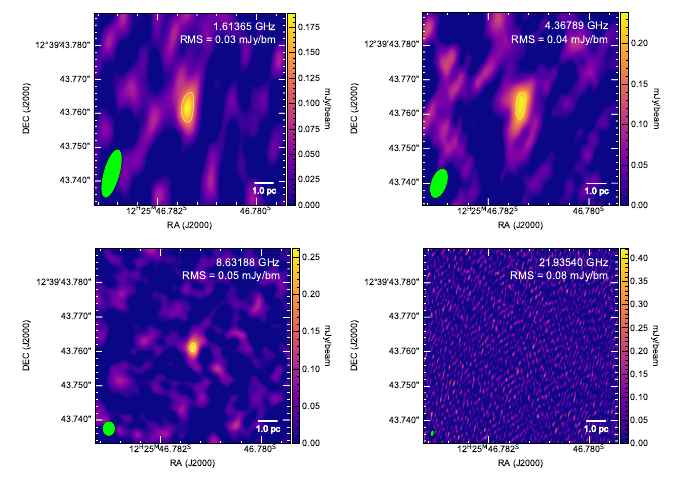}
\includegraphics[clip,trim=0.2cm 0cm 0cm 0.3cm, width=0.3\textwidth]{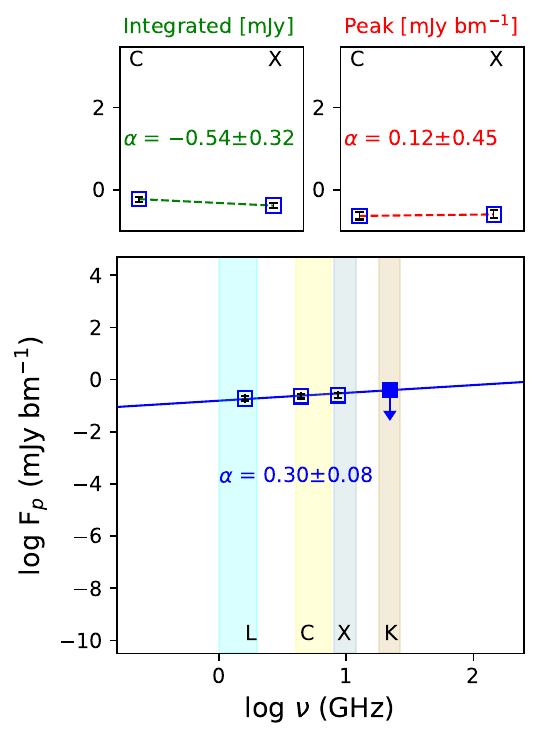}\\
\vspace{1cm}
\large \textbf{NGC 4593}\par\medskip
\includegraphics[clip, trim=5cm 15.3cm 5cm 4.3cm, width=0.67\textwidth]{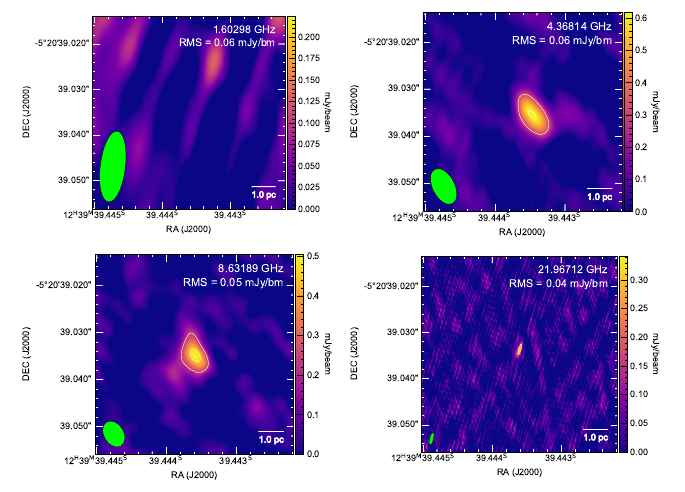}
\includegraphics[clip,trim=0.2cm 0cm 0cm 0.3cm, width=0.3\textwidth]{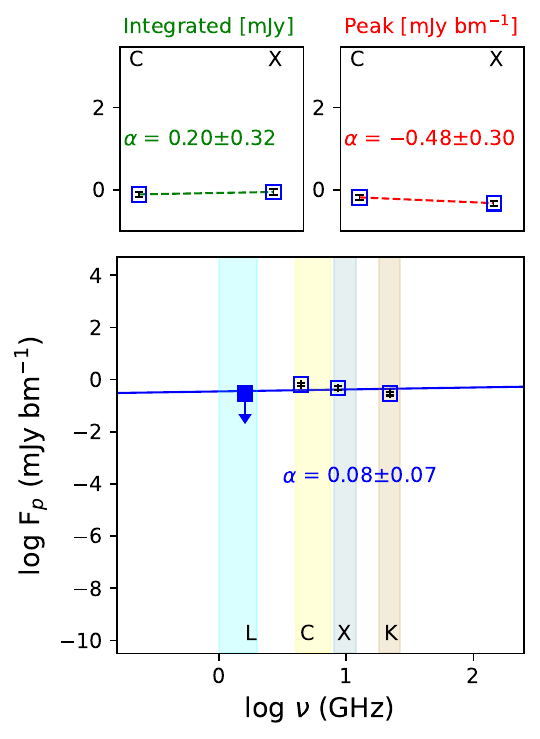}\\
\caption{}
\end{figure*}

\addtocounter{figure}{-1}

\begin{figure*}[htp]
\centering
\vspace{1cm}
\large \textbf{NGC 5290}\par\medskip
\includegraphics[clip, trim=5cm 15.3cm 5cm 4.3cm, width=0.67\textwidth]{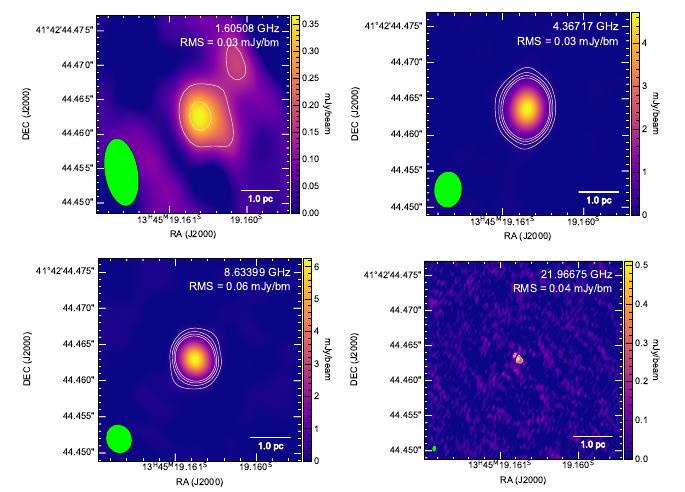}
\includegraphics[clip,trim=0.2cm 0cm 0cm 0.3cm, width=0.3\textwidth]{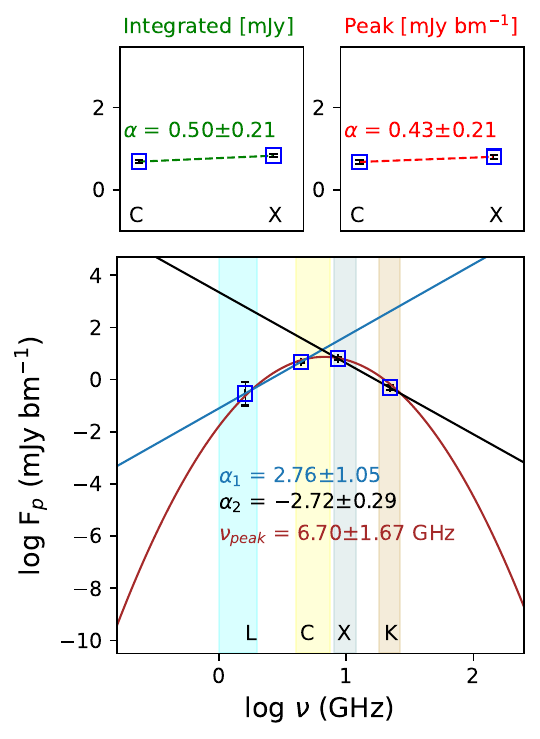}\\
\vspace{1cm}
\large \textbf{NGC 5506}\par\medskip
\includegraphics[clip, trim=5cm 15.3cm 5cm 4.3cm, width=0.67\textwidth]{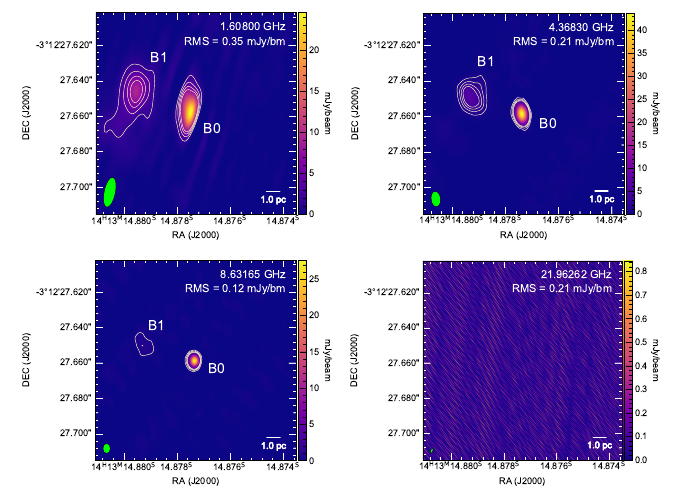}
\includegraphics[clip,trim=0.2cm 0cm 0cm 0.3cm, width=0.3\textwidth]{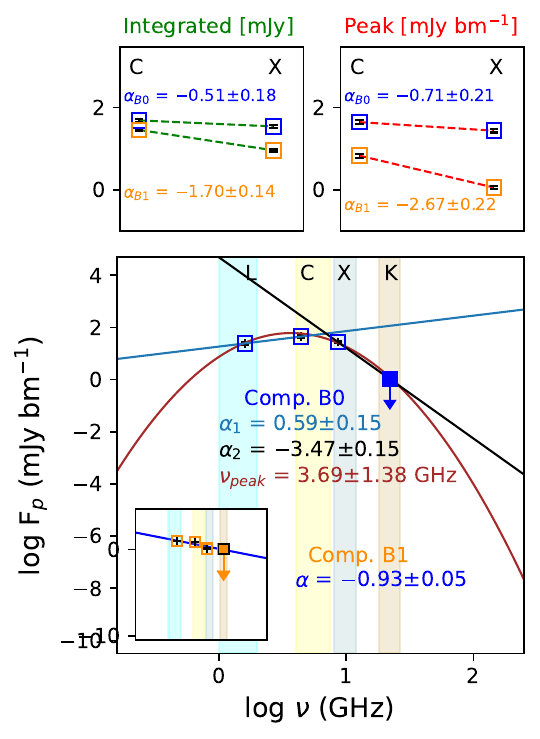}\\
\caption{}
\end{figure*}

\noindent{\textbf{NGC 3079 [10$^{\rm h}$01$^{\rm m}$57$\fs$804, $+$55$\degr$40$\arcmin$47$\farcs$239]}}: NGC 3079 is one of the two sources in our sample with multicomponent (A and B; mentioned in~\citet{trotter_1998ApJ...495..740T}) radio sources. In addition, a third component lying between A
and B was tentatively identified in C and X bands, but its characteristics were
not well constrained by the observation.  Source A was detected in L, C, and X bands, whereas component B was seen only in C and X bands. Including non-detections upper limits (5$\sigma$), both components showed a GHz peaked spectrum and slightly resolved point-like structures. Component A showed a turnover peak at 4.46 GHz, and B showed a turnover peak at 5.60 GHz. After the turnover frequency, both components showed comparatively steeper spectra at high frequencies. Component A's turnover frequency is lower than B, so component A showed a decrement of peak intensity starting near the C band. In contrast, component B's peak intensity values keep rising till the X band (as the turnover peaks at a higher frequency) with an inverted spectrum and then drop steeply at higher frequencies. This difference is also seen in the two additional SED plots using only C and X band peak intensities and integrated flux densities. \\

\noindent{\textbf{NGC 3516 [11$^{\rm h}$06$^{\rm m}$47$\fs$46352, $+$72$\degr$34$\arcmin$07$\farcs$2784]}}: A flat spectrum ($\alpha$ $> -$0.5 ) was found for NGC 3516 with L, C, and X band detections and K band upper limit. SEDs from the peak intensities and integrated flux densities in C and X bands show flat spectra and suggest a point source. \\

\noindent{\textbf{NGC 4151 [12$^{\rm h}$10$^{\rm m}$32$\fs$58201, $+$39$\degr$24$\arcmin$21$\farcs$063]}}: All four band images revealed a bright central sub-parsec scale emission with a ``candle flame'' shape outflow in the northeast direction. The elongated structure is visible in all four band images, even in high-resolution, K band sub-parsec ($\sim$0.1 pc) spatial scale.  The interesting jet/outflow structures of NGC 4151 were previously reported in various sub-parsec to parsec scale VLBI and e-MERLIN studies~\citep{Ulvestad_2005AJ....130..936U, Williams_2017MNRAS.472.3842W, Williams_2020MNRAS.495.3079W}.  While plotting SEDs using only C and X band peak intensities and integrated flux densities, we found flat to slightly steeper spectra, suggesting that the central region is a compact point source. Four frequency SED model using all peak intensities showed a GHZ peaked spectrum with flatter in lower frequency regime and a steeper spectrum in higher frequency as K band peak emission is relatively weak.\\ 

\noindent{\textbf{NGC 4235 [12$^{\rm h}$17$^{\rm m}$09$\fs$88175, $+$07$\degr$11$\arcmin$29$\farcs$670]}}: L, C, and X band images showed a point-like structure, with a little extended feature in the X band image; thus, the integrated flux density in X band is a little higher compared to C band flux density. NGC 4235 is another example of our sample with a GHz peaked spectrum found in four-frequency SED with a turnover frequency of about $\sim$4.3 GHz. Higher frequency spectrum, including the 5$\sigma$ nondetection upper limit in the K band, showed a steeper spectral index ($\alpha \approx -$1.6).  \\ 

\noindent{\textbf{NGC 4388 [12$^{\rm h}$25$^{\rm m}$46$\fs$78135, $+$12$\degr$39$\arcmin$43$\farcs$761]}}: All four band images revealed a weak central radio emission. NGC 4388 is the weakest of all radio sources detected in the L, C, and X bands. No K-band emission was detected.  The four-frequency SED plot with the K band upper limit shows an inverted spectrum with a spectral index of $\sim$0.3.\\

\noindent{\textbf{NGC 4593 [12$^{\rm h}$39$^{\rm m}$39$\fs$44359, $-$05$\degr$20$\arcmin$39$\farcs$035]}}: Similar to NGC 4388, images reveal a weak compact core-like emission at the center. The four-frequency SED plot shows a flat spectrum with $\alpha = $0.08 with L-band upper limit.  \\

\noindent{\textbf{NGC 5290 [13$^{\rm h}$45$^{\rm m}$19$\fs$16076, $+$41$\degr$42$\arcmin$44$\farcs$463]}}: All four band images show a bright point source at the center. The SED sub-plots using only C and X bands show a similar flat spectrum, supporting the idea of a point source. The four-frequency SED plot shows a GHz-peaked spectrum with a peak at 6.70 GHz. A similar radio peak intensity drop was seen at lower and higher frequency ends. \\ 

\noindent{\textbf{NGC 5506 [14$^{\rm h}$13$^{\rm m}$14$\fs$87735, $-$03$\degr$12$\arcmin$27$\farcs$6585]}}: We detected two components (B0 and B1; mentioned in~\citet{Middelberg_2004A&A...417..925M}) in L, C, and X band images. Neither component was detected in the K band. All three detected images and central two-frequency (C and X bands) SEDs using peak intensities and integrated flux densities suggested a pair of two point-like sources slightly resolved in the C band. But their four-frequency SED plot shows that component B0 follows a GHz-peaked spectrum with a turnover peak near $\sim$4 GHz, whereas component B1 peak intensities dropped with the frequency increase resulting in a steeper spectrum.\\

Table~\ref{tab:alpha} lists all our sources' spectral shapes and indices, including turnover frequencies from the GHz-peaked spectrum, lower and higher frequency domain spectral indices ($\alpha_{1}$ and $\alpha_{2}$) and spectral indices derived from C and X band peak intensities and integrated flux densities only ($\alpha^{int}$ and $\alpha^{peak}$).

\section{Discussion} \label{sec:dis}

\subsection{Detection Rate and Central Radio Emission}

In FRAMEx Papers I and III, with the help of VLBA sub-parsec scale ($\sim$milli-arcsec resolution) observations in the C (5.8 GHz) band, the detection fraction was $\sim$33-36\%, in contrast with lower resolution kpc scale VLA (archival)  observations ($\sim$ arcsec resolution), taken in A configurations at $\sim$ 4.89 GHz, where the detection fraction was about $\sim$100\%. We suggested that the high detection fraction in the kpc scale is due to the extra-nuclear radio emission from interactions between the AGN and host environment. 
Now in this work, we observed 12 detected sources from Paper I and III in L (1.6 GHz), C (4.4 GHz), X (8.6 GHz), and K (22 GHz) bands using VLBA with a sensitivity depth of $\sim$25 $\mu$Jy bm$^{-1}$ per band to probe the actual origin of radio emission in the sub-parsec regime.  

We detected 12/12 of our sources in the C and X bands, whereas 10/12 ($\sim$83\%) sources were detected in L and 6/12 ($\sim$50\%) in the K band. We found a relatively higher detection rate than previous high-resolution (VLBA, eMERLIN) studies at low-frequency (L band/1.6 GHz). For example, in a study of 103
nearby galaxies from the Palomar sample observed at 1.5 GHz with the high-resolution ($\sim$150 mas) eMERLIN array,  they found a $\sim$50 percent of detection rate~\citep{Baldi_2018MNRAS.476.3478B}. A similar fraction of detection was reported in a VLBA study of a sample of radio-loud(RL) and radio-quiet(RQ) AGNs at 1.4 GHz in \citet{Maini_2016A&A...589L...3M}. In addition, all these studies found that most detected sources are either single radio core detection or sometimes associated with more complex structures similar to ours. The VLBA detected sub-parsec scale emission at 1.6 GHz in 83 $\%$ of our targets. In FRAMEx Paper II, we showed that synchrotron self-absorption could cause low detection rates at VLBA spatial scales at C band frequencies. Thus, at the L band, one would infer these effects to be more prominent as synchrotron self-absorption is even more dominant at 1.6 GHz than at 5-6 GHz. 

At the higher end of the spectrum, we found 50 percent detection with the VLBA in the K-band. In contrast, in a  JVLA study of 100 ultra-hard X-ray selected AGN at K band (22 GHz), \citet{Smith_2020MNRAS.492.4216S} detected 96 percent of their sample. They found a compact radio core detection for all sources. But their comparatively low-resolution ($\sim$1 arcsec) radio study did not differentiate between synchrotron emission from AGN cores and thermal or free-free emission from dust or star formation. Though our detection fraction is not as high as this previous study, the higher-resolution observations, including the K band, found the radio-emitting regions within a parsec of the center engine of the AGNs. A low detection rate or lower luminosity in the K band at pc-scale suggests the synchrotron emission can still be detected in higher frequency (until or beyond 22 GHz) in opposition to large-scale VLA radio detections where the thermal emission from dust can dominate in that frequency range.   

Now the question that needs to be addressed is whether this central radio emission is associated with collimated outflows (for example, small-scale jets), accretion disk corona, or interaction between nearby ISM and small-scale AGN outflows or winds. In the next section, we investigate different physical mechanisms at work in our sample of AGNs using sub-parsec scale radio spectral index analysis.


\subsection{Giga-Hertz Peaked Spectrum (GPS)} \label{sec:GPS}

\renewcommand{\thefigure}{\arabic{figure}}
\begin{figure}
\includegraphics[width=\columnwidth]{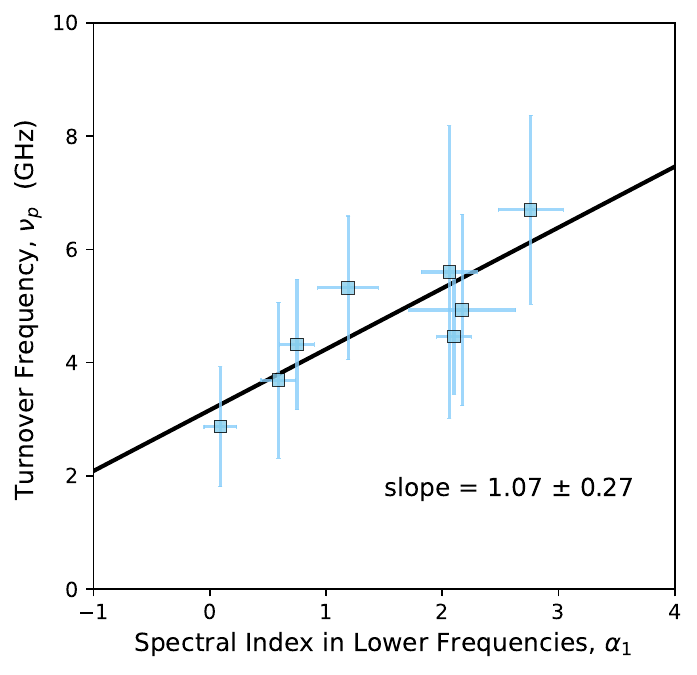}
\caption{Correlation between the inverted spectral index at lower frequencies and turnover (peak) frequencies in GPS sources. The peak frequencies have large uncertainties due to the contribution of L and K band nondetection upper limits.}
\label{fig:alpha1_peak}
\end{figure}

Giga-Hertz peaked spectrum (GPS) is a term used to describe a type of radio source that emits radiation predominantly in the GHz frequency range, with a peak (turnover) in its SED at GHz frequencies (typically $\sim$1 GHz; \citet{de_vries_1997A&A...321..105D}). 
The turnover frequency is characterized by a change in the spectral index (the slope of the SED) from a flatter value at lower frequencies to a steeper value at higher frequencies.  At lower frequencies, the radiation can be absorbed by the relativistic electrons (synchrotron self-absorption) and/or by thermal electrons (free-free absorption), causing a flat/inverted slope in the low-frequency domain. Higher energy electrons can radiate their energy, causing steeper spectra at high frequencies. These phenomena combined create convex-shaped radio spectra with a spectral turnover frequency between hundreds of MHz and a few GHz~\citep{snellen_1998A&AS..131..435S, Patil_2018ASPC..517..595P}. The physical mechanisms responsible for the GPS phenomenon are still poorly understood, but they are believed to be young, compact radio sources representing the early stages of massive RL AGNs' life cycles~\citep{Fanti_1990A&A...231..333F, Odea_1991ApJ...380...66O, Odea_1998PASP..110..493O,
lister_2003ASPC..300...71L, Sotnikova_2019AstBu..74..348S,patil_2020ApJ...896...18P}.

A recent study of high-frequency, high-resolution VLBI observations of GPS sources by~\citet{Cheng_2023arXiv230303066C} mainly found a compact core (associated with the one-sided jet feature) and a turnover frequency ranging from 6 and 32 GHz. Similarly, our study found GPS sources (7 out of 12) mainly with parsec to sub-parsec scale central radio-emitting region, few with extended outflow-like structure (e.g., NGC 1052), and multicomponent features (e.g., NGC 3079, NGC 5506). The turnover frequencies range from $\sim$3-6 GHz (see Table~\ref{tab:alpha}). The turnover scenario is consistent with the idea we proposed in Paper III, free–free or self-absorbed synchrotron radiation or a combination of both can happen at lower frequencies ($\leq$ 5 GHz), where shocks accelerate relativistic electrons within the accretion flow of AGNs. In addition, a model proposed by~\citet[]{Ishibashi_2011A&A...525A.118I} suggested that the opacity decreases with the frequency increase and the transition from optically thick to optically thin regimes lies around a few tens of GHz. So, in the low-frequency regime for VLBA resolutions, we can expect inverted spectra due to the radiation absorbed by the relativistic electrons, which would lead to a peak near a few GHz. 

We found an interesting linear correlation, $\nu_{p} = (1.07\pm0.27)*\alpha_{1} + 3.16$, when we plotted turnover frequencies ($\nu_{p}$) with lower frequency spectral indices or $\alpha_{1}$ (Figure~\ref{fig:alpha1_peak}). This correlation means that the radio intensity drops more rapidly or, in other words, absorption is higher (free-free or synchrotron) at lower frequencies for the sources with higher turnover frequencies.  Recent proper motion studies and multifrequency polarization observations established that GPS sources are young objects embedded in a dense interstellar medium and interacting with it~\citep{Odea_2021A&ARv..29....3O}. Previously from the observational studies of GPS sources, an anti-correlation was found between turnover frequency and source size~\citep{odea_1997AJ....113..148O}. That means the youngest objects have the highest turnover frequencies, and
as the source expands and the energy density
decreases, the peak frequency is expected to move toward a lower
frequency~\citep{tinti_2005A&A...432...31T}. These results combined established that the absorption is higher in the dense central region of young radio AGNs. As most of our sources are relatively higher accreting AGNs (see Section~\ref{accretion}), higher density is expected in their central parsec regions, representing the young radio AGN group.

\begin{deluxetable*}{l|ccc|cc}[ht!]
\label{tab:alpha}
\tablecaption{Radio Spectral Index} 
\tablehead{\textbf{GHz peaked}& $\nu_{peak}$& $\alpha_{1}$ & $\alpha_{2}$& $\alpha_\mathrm{(4.4-8.6)}^{int}$&$\alpha_\mathrm{(4.4-8.6)}^{peak}$\\
& (1)& (2) & (3)& (4)&(5)}
\startdata
NGC 1052 & 5.32 $\pm$ 1.27 & 1.19 $\pm$ 0.26& $-$2.27 $\pm$ 0.18 &~~0.36 $\pm$ 0.41&~~0.53 $\pm$ 0.34\\
NGC 2110 & 4.93 $\pm$ 1.69& 2.17 $\pm$ 0.46& $-$5.06 $\pm$ 0.26& $-$0.28 $\pm$ 0.19&$-$0.51 $\pm$ 0.21\\
NGC 3079 (A) & 4.46 $\pm$ 1.02& 2.10 $\pm$ 0.15& $-$4.62 $\pm$ 0.15&$-$0.86 $\pm$ 0.14&$-$1.25 $\pm$ 0.21\\
NGC 3079 (B) & 5.60 $\pm$ 2.58&2.06 $\pm$ 0.24& $-$4.84 $\pm$ 0.15 &~~1.46 $\pm$ 0.61&~~1.37 $\pm$ 0.21\\
NGC 4151 & 2.87 $\pm$ 1.06 & 0.09 $\pm$ 0.14&$-$3.07 $\pm$ 1.22&$-$0.40 $\pm$ 0.17&$-$0.89 $\pm$ 0.21\\ 
NGC 4235 & 4.32 $\pm$ 1.15& 0.75 $\pm$ 0.15&$-$1.57 $\pm$ 0.16&~~0.17 $\pm$ 0.24&$-$0.65 $\pm$ 0.24\\
NGC 5290 & 6.43 $\pm$ 1.32& 2.76 $\pm$ 0.28& $-$2.72 $\pm$ 0.22&~~0.50 $\pm$ 0.21&~~0.43 $\pm$ 0.21\\  
NGC 5506 (B0) & 3.69 $\pm$ 1.38& 0.59 $\pm$ 0.15& $-$3.47 $\pm$ 0.15&$-$0.51 $\pm$ 0.18&$-$0.71 $\pm$ 0.21\\
\hline
\textbf{Steep/flat} &  \multicolumn{3}{c|}{$\alpha_\mathrm{(1.6-22)}$}&&\\
\textbf{/inverted}&  \multicolumn{3}{c|}{(6)}&&\\
\hline
NGC 1068 &\multicolumn{3}{c|}{$-$0.41 $\pm$ 0.05}   & $-$2.83 $\pm$ 0.27& $-$0.99 $\pm$ 0.28\\
NGC 2992 & \multicolumn{3}{c|}{$-$0.35 $\pm$ 0.07}     &$-$0.43 $\pm$ 0.31&$-$0.20 $\pm$ 0.25\\
NGC 3516 &\multicolumn{3}{c|}{$-$0.31 $\pm$ 0.08}      &~~0.19 $\pm$ 0.23&$-$0.09 $\pm$ 0.28\\
NGC 4388 &\multicolumn{3}{c|}{~~0.30 $\pm$ 0.08}      &$-$0.54 $\pm$ 0.32& ~~0.12 $\pm$ 0.45\\
NGC 4593 &   \multicolumn{3}{c|}{~~0.08 $\pm$ 0.07}   & ~~0.20 $\pm$ 0.32& $-$0.48 $\pm$ 0.30\\
NGC 5506 (B1) & \multicolumn{3}{c|}{$-$0.93 $\pm$ 0.05}  &$-$1.70 $\pm$ 0.14 &$-$2.67 $\pm$ 0.22\\
\enddata  
\tablecomments{ (1) turnover frequency from best-fitting quadratic polynomial function. \\ 
(2) $\&$ (3) $\alpha_{1}$ and $\alpha_{2}$ were measured using
piecewise linear regression fit for the sources with giga-hertz peaked spectrum.\\
(4) $\&$ (5) $\alpha_\mathrm{(4.4-8.6)}^{int}$ and $\alpha_\mathrm{(4.4-8.6)}^{peak}$ were measured between 4.4 and 8.6 GHz using their respective peak intensities and integrated flux densities derived from CASA’s 2-D Gaussian model fitting algorithm.\\
(6) $\alpha_\mathrm{(1.6-22)}$ values were measured for the sources with a single power-law spectrum between 1.6 and 22 GHz.}
\end{deluxetable*}

\subsection{Steep, Flat or Inverted Spectra} \label{sec:steep/flat}
~\citet{hovatta_2014AJ....147..143H} showed differences in distributions of observed spectral indices between core and jet components of AGN from a four frequency (8.1, 8.4, 12.1, and 15.4 GHz) VLBA observation of 190 extragalactic radio jets in the MOJAVE survey. They demonstrated that the emission with a flatter ($\alpha$ $>$ $-$0.5)~
spectrum is generated in the inner part or close to the central core, and jet spectra are relatively steeper with a mean index value $\alpha$ $\approx$ $-$1. More recently, in a (sub)-kpc  Very Large Array (VLA) study of a sample of 30 nearby (RQ) AGN, ~\cite{panessa_2022MNRAS.515..473P} found $\sim70\%$ of the sources either with steep spectra ($\alpha$ $\leq -$0.5), or flat spectra ($\alpha$ $> -$0.5). They suggested that the steep spectra are compatible with optically-thin synchrotron from a jet or disk winds, which are unresolved in compact cores, and flat spectrum sources are generally compact, indicating a possible optically-thick synchrotron emission from a compact jet, a hot corona, or both. They also calculated the mean of the slopes of the cores and found a nearly flat gradient ($\alpha$ $> -$0.5) had been observed from 5 to 15 GHz, which steepened sharply from 15 to 22 GHz ($\alpha$ $< -$1.0). So, optically thick core synchrotron emission was suggested below 15 GHz for their sample of relatively higher accreting RQ AGNs, which becomes optically thin above 15 GHz. Now, if any small (parsec scale) optically thin radio lobe/jet was the possible physical mechanism here, then the slope might have been steeper in lower frequencies as well~\citep{Gultekin_10.1093/mnras/stac2608}. 

In principle, steeper spectra are associated with optically-thin synchrotron jets (unresolved in compact cores). A flatter spectrum is associated with optically-thick inner core radio emission from winds or shocks. We found the remaining 5/12 source with nearly flat or inverted spectra ($\alpha$ $> -$0.5 or $\alpha$ $>$ 0 ) and a component of NGC 5506 (B1) separated by 4--5 pc from the central B0 source showed steep spectra with $\alpha$ $< -$0.5 (see Table~\ref{tab:alpha}). Sources with flat spectra showed mostly resolved/slightly resolved bright central regions establishing that the radio emission is from pc-scale winds/shocks at the immediate vicinity of these AGNs. The NGC 5506 (B1) component with a steep spectrum ($\alpha$ $\sim -$0.93) may originate from a small-scale optically thin, one-sided, strong collimated outflow or interaction with neighboring ISM. On the other hand, the spectral indices of NGC 1068, NGC 2992, and NGC 3516 were very similar, indicating a flat ($\alpha > -$0.5) spectrum. So, their radio emission might be due to unresolved sub-parsec scale winds. NGC 4593 and NGC 4388 might share a similar origin of radio emission as they showed nearly inverted to inverted spectra. ($\alpha$ $\geq$ 0). In addition, these two sources showed the weakest radio emission compared to other sources in our sample. The central environment, e.g., accretion and star formation, plays a crucial role in determining the true nature of radio emission on such a small spatial scale.

\begin{figure*}
\centering
\includegraphics[scale = 0.8]{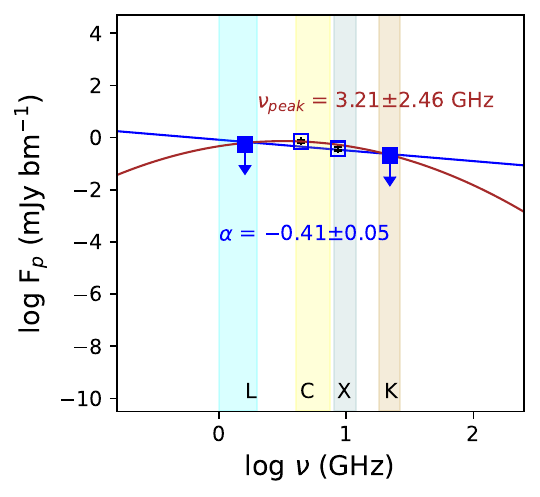}
\includegraphics[scale = 0.8]{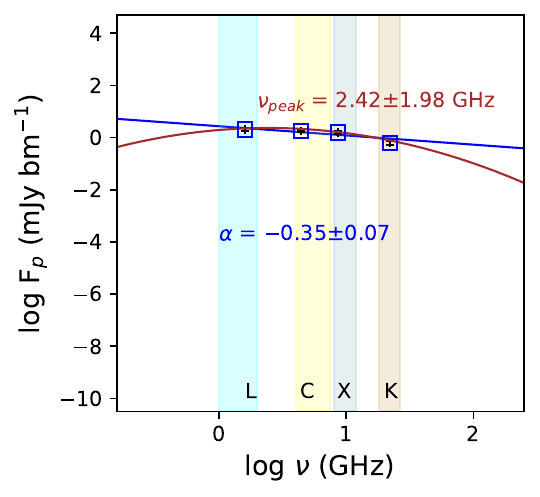}
\caption{NGC 1068 (left) and NGC 2992 (right) are examples of nearly steep spectra sources or may be peaked spectrum sources with lower turnover frequency than other GPS sources in our sample. It is believed that lower turnover frequency sources might be in the transient or intermittent stage between GPS sources (few giga-hertz) to CSS sources (hundreds of mega-hertz).} 
\label{fig:CSS}
\end{figure*}

In the next few sections, we will discuss other alternate possibilities of an AGN having a steep or flat spectrum. They may be part of the bigger picture of AGN evolutionary stages or completely different AGN classes from GPS sources.

\subsection{Evolutionary Stages: From GPS to CSS} \label{sec:CSS}
Possible scenarios for the sources with nearly steep or steep spectra would be that they are GPS sources with lower turnover frequency, or they might be in the evolutionary stage and continue to grow from GPS to compact steep spectrum (CSS) sources, where their turnover frequency will decrease to few hundreds of MHz. It is believed that the GPS sources are in the earliest stage of evolution or young radio galaxies that will eventually evolve into CSS sources on their way to becoming large radio galaxies~\citep{an_2012ApJ...760...77A,Odea_2021A&ARv..29....3O}. CSS sources tend to have sizes between 500 pc and 20 kpc, while GPS sources have projected linear sizes of less than 500 pc~\citep{Odea_2021A&ARv..29....3O}. Figure 3 of the FRAMEx paper I showed C-band (4.9 GHz) hundred-parsec-scale radio morphologies from VLA observations, where we can see that NGC 1068, NGC 2992, and NGC 5506 (nearly steep/steep spectrum) have hundreds of kpc scale extended radio emission (either lobe-like or diffuse structure) compared to the sources with giga-hertz peaked spectrum, which show much more collimated small-scale outflow features (for example, NGC 1052, NGC 3079, NGC 4151). In Figure~\ref{fig:CSS} (similar to what we have shown in Figure~\ref{fig:multiband_image}, but with curved power-law fit), we showed that NGC 1068 and NGC 2992 are examples of two possible low turnover frequency GPS sources that might be transient or intermittent in the evolutionary phase and will finally become a CSS source.  


But a low turnover frequency indicates that NGC 1068 or NGC 2992 are older than sources like NGC 1052 or NGC 2110, whereas the opposite is true (see Section~\ref{accretion} for details). NGC 1068 has a much higher accretion rate than NCG 1052, making the central environment of NGC 1068 more chaotic and representing a younger AGN population. Though the accretion rate and our study here alone can not differentiate the actual age of the AGNs in our sample, we believe that the nearly steep or flat/inverted 
spectra sources are examples of different classifications of AGNs (radio-quiet) rather than transient or specific stages of evolution between different aged AGNs.


\subsection{Accretion rate and Radio Spectra}\label{accretion}

\renewcommand{\thefigure}{\arabic{figure}}
\begin{figure*}
\centering
\includegraphics[width=\columnwidth]{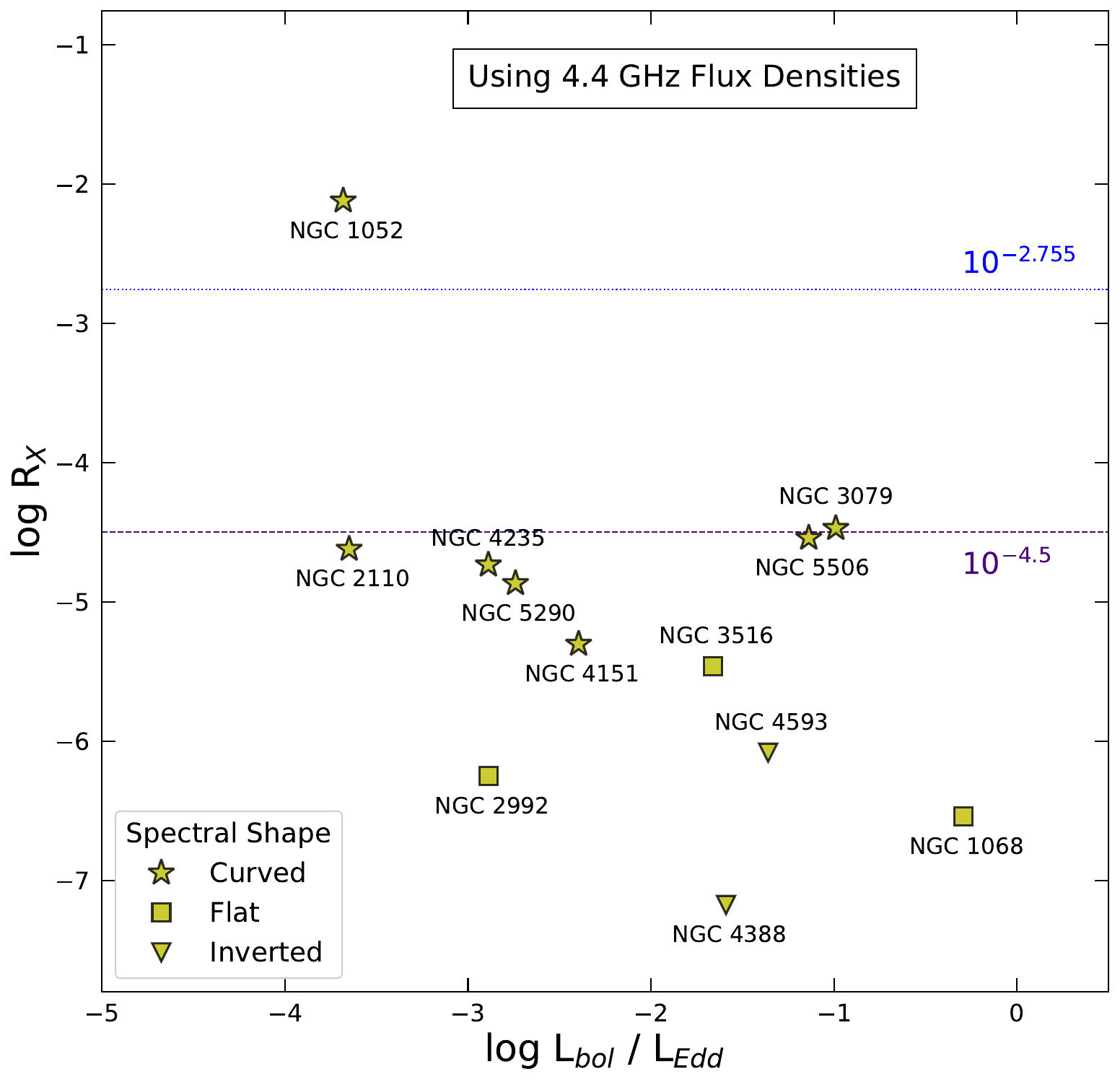}
\hspace{2mm}
\includegraphics[width=\columnwidth]{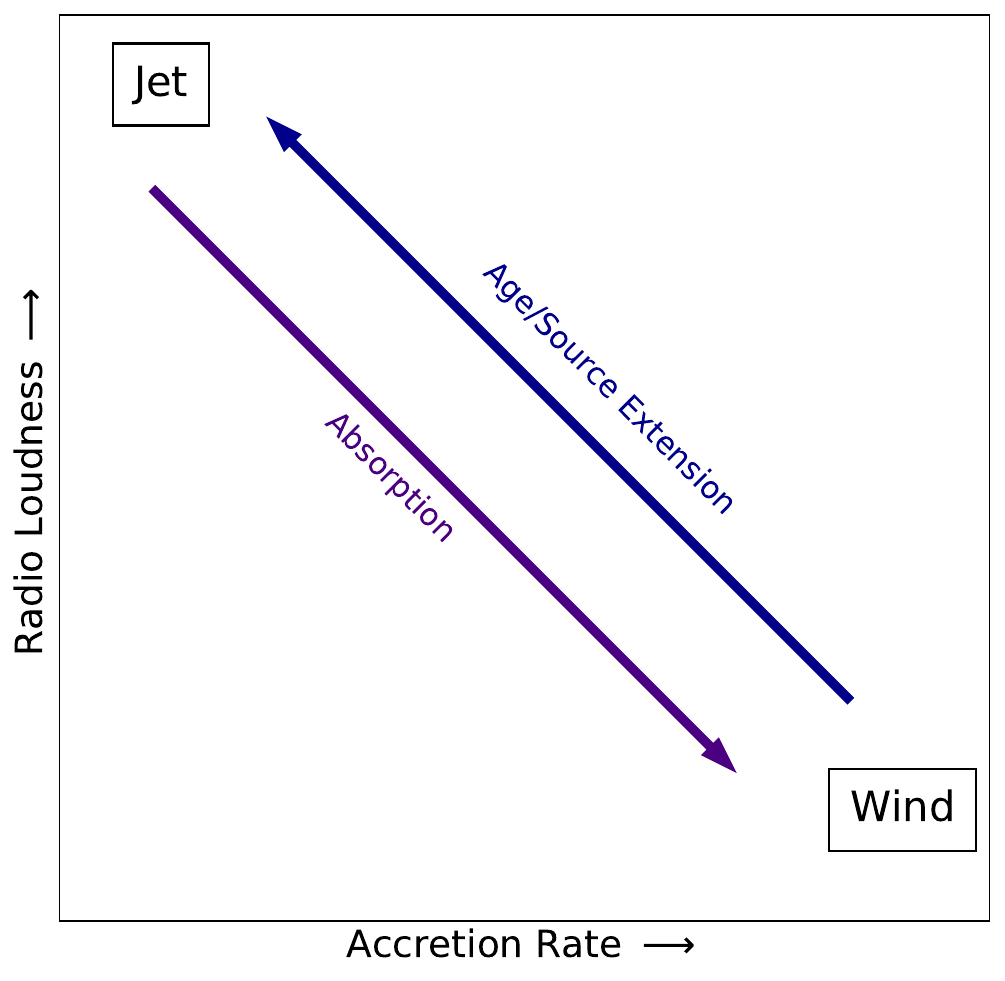}
\caption{(\textit{left:}) Radio loudness vs. Eddington ratio plot for our sample with their spectral shapes, where radio loudness was determined using flux densities from the 5$\sigma$ contour region of our 4.4 GHz observations. The blue dotted line indicates the separation between radio-loud (RL) and radio-quiet (RQ) AGNs as introduced by \citet{panessa_2007A&A...467..519P} and separates Seyferts from LLRGs of $L_\mathrm{R}$/$L_\mathrm{X}$ = 10$^{-2.755}$. The purple dashed line represents the classical division between RL and RQ AGNs by~\citet{Terashima_2003ApJ...583..145T}. The plot shows that the flat and inverted spectra are associated with the sources with higher accretion rates and lower radio luminosities. (\textit{right:}) a schematic diagram representing differences in physical mechanisms responsible for radio emission as a function of absorption and source age in the same parameter space.} 
\label{fig:Rx_vs_accretion}
\end{figure*}
In Figure 7 of Paper III, we showed that the radio loudness parameter,~$R_\mathrm{X} = L_\mathrm{6~cm} / L_\mathrm{2-10 keV}$, follows a negative trend when plotted as a function of the Eddington ratio $L_{bol}/L_{Edd}$, 
which parametrizes the source's accretion rate. 
Here we plotted a similar figure (\textit{left} panel on Figure~\ref{fig:Rx_vs_accretion}), but this time, we used 4.4 GHz core flux densities (total flux density within 5$\sigma$ contour) instead of 6 GHz (6 cm) peak intensities to determine the radio loudness. We also showed radio spectral shape for each source we observed in this study. We find that sources with flat/inverted spectra are grouped in the region of highest accretion and lowest radio loudness, bottom right corner of the plot. The high accretion regime in black hole systems favors wind-based radio emission, similar to different accretion states observed in XRBs in their transition from the ``low/hard'' state associated with steady radio emission attributed to a jet to the ``high/soft" state where the radio emission is suppressed. This wind-jet anti-correlation scenario has been discussed before in \citet[][]{Wang_2004ApJ...615L...9W,Sikora_2007ApJ...658..815S,Mehdipur_2022ApJ...925...84M}. This trend also appeared when we plotted the radio loudness parameter as ~$R_\mathrm{X} = L_\mathrm{8.6 GHz} / L_\mathrm{2-10 keV}$, where 8.6 GHz luminosities were measured from the total radio flux densities (5$\sigma$) of our observations (Figure~\ref{fig:Rx_vs_accretion_appendix} in Appendix). 
If the wind is strong enough to produce shocks or outflows, the radio emission will result from the interaction between outflows and the neighboring ISM in the central region.  If the wind is not strong enough or there is no ISM-wind interaction, then the radio emission from sources near or above the log $R_{X}$ = $-$4.5 line by \citet{Terashima_2003ApJ...583..145T}, which represents the traditional separation between RL and RQ AGNs (which corresponds to a value of 10 for the optically-based radio-loudness parameter R), originate either from coronal emission or have a jet. 

Based on this wind-jet scenario, we make a schematic model of the area where different physical mechanisms can dominate to produce radio emission (\textit{right} panel in Figure~\ref{fig:Rx_vs_accretion}). We previously showed in Figure 8 of Paper III that the low radio luminosity of an AGN is due to a combination of the synchrotron self-absorption and the presence of wind/shocks. Here we showed in our schematic that the obscuration or absorption increases for the sources with the highest accretion and lowest radio-loudness (purple arrow). On the other hand, the anti-correlation between the source extension and turnover frequency discussed in Section~\ref{sec:GPS}, as well as the conclusion from the discussion of young radio sources with relatively higher accretion rates, allowed us to demonstrate that the extended collimated outflow/larger source size is associated with the lowest accretion and highest radio-loudness values (blue arrow).

\subsection{Star Formation in Sub-pc Scale}\label{star_formation}

In a study of 62 RQ AGNs with arcsec resolution,~\citet{Smith_2016ApJ...832..163S} showed that radio emission (especially for sources with extended emission) in lower resolution ($\sim 1 \arcsec$ VLA) imaging surveys might not be associated with AGN core synchrotron emission but with star formation. A large number of previous studies of the closest Seyfert 2 galaxies suggested a link (but no direct connection) between AGNs and young star formation on a scale of a few hundred parsecs~\citep[and references therein]{Davies_2007ApJ...671.1388D}{}. Typically the connection between these two phenomena is discussed from kpc to a few hundred-parsec scales. Now, to find evidence of a stellar population in the closest proximity of AGN (few pc to sub-pc scale), one approach~\citep{Middelberg_2004A&A...417..925M} is to measure the proper motion of radio components close to the nucleus using multiple
epochs of observations with Very Long Baseline Interferometry (VLBI). \citet{Middelberg_2004A&A...417..925M} discussed evidence of non-relativistic nuclear compact radio components found in many Seyferts, mostly sources with low radio luminosities.  They also suggested flat-spectrum core components for the powerful radio galaxies with a parsec-scale resolution while analyzing spectral indices. We found similar flat/inverted spectrum components but associated with weak central radio-emitting sources (NGC 1068, NGC 2992, NGC 4388, NGC 4593). These sources showed no sign of powerful, collimated, and extended features, only indicating a compact central radio emission, possibly harboring non-relativistic star-forming components. 

\begin{figure}
\includegraphics[width=\columnwidth]{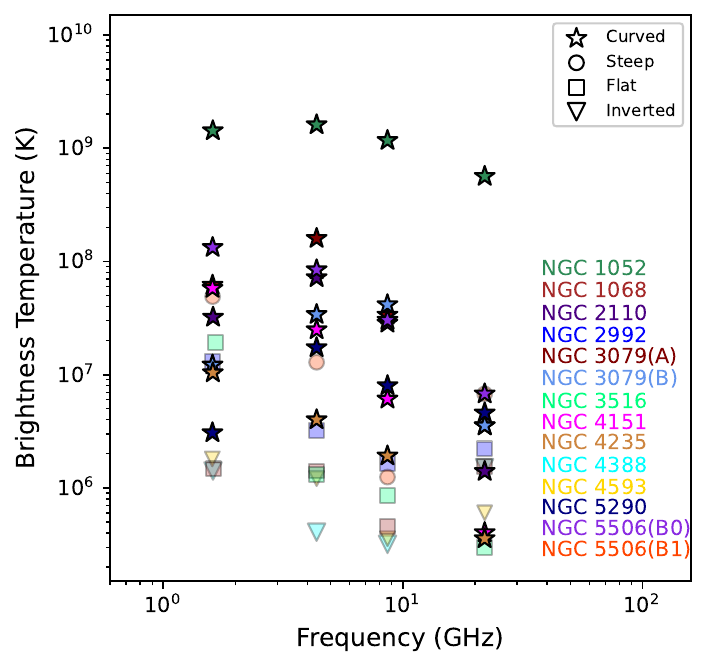}
\caption{Brightness temperature (K) as a function of observed frequency (GHz). The figure shows measurements of brightness temperature using the peak intensities at 1.6 GHz, 4.4 GHz, 8.6 GHz, and 22 GHz. The brightness temperature of sources with GHz peaked, or curved spectra lies well above that of sources with flat/inverted spectra in all frequencies.}
\label{fig:tb_freq}
\end{figure}

In addition, to extend our star formation analysis and to find any contribution from star formation to the nuclear radio emission of our sources, we calculated radio-based (we used 1.6 GHz instead of 1.4 GHz) star formation rate (SFR) from~\citet[equation 6]{Herrero_2017MNRAS.471.1634H}{}, which is $\mathrm{SFR} = 1.02\times10^{-28} \mathrm{L_{1.4 GHz}}$.  The approach for determining SFR using radio data was proposed by~\citet{Murphy_2011ApJ...737...67M}  and relies on the assumption of a strong correlation between IR and radio luminosities. However, the radio luminosity at this frequency for sources with flat/inverted spectra is lower than most GHz peaked sources, and so is the calculated SFR. Radio emission at lower frequencies is essentially due to non-thermal emission, and we cannot distinguish the thermal contribution using this method of calculating SFR. On the other hand, in Figure~\ref{fig:tb_freq}, we presented a plot illustrating the relationship between brightness temperature ($T_b$) and rest-frame frequency. Our findings indicate that sources with flat/inverted spectra exhibit lower $T_b$ than those with steep and GHz peaked spectra. Specifically, the $T_b$ values for sources with flat/inverted spectra range between $10^{6.5}$~K to $10^{5.6}$~K (in C band). This suggests that thermal free-free emission could mainly contribute to radio emission. 

\section{Conclusions} \label{sec:conclusions} 

We present results from high-resolution multi-wavelength radio observations of 12 detected AGNs from our previous campaigns to look into the sub-parsec regime of AGNs and study their core radio emission. We obtained VLBA observations at L ($\sim$1.6 GHz), C ($\sim$4.4 GHz), X ($\sim$8.6 GHz), and K ($\sim$22 GHz) radio bands with homogeneous observing setups. The primary goal of this paper was to determine the ``True Central Origin of Radio Emission (TCORE)'', and disentangle contributions from potentially multiple sources and different physical mechanisms by spectral index analysis. Our main findings are listed below.
{\begin{enumerate}
    \item We found a relatively higher detection rate ($\sim$83\%) than previous high-resolution radio studies at low-frequency (L band). All of our sources were detected in the C and X bands. On the other hand, from high-frequency K-band observations, we were able to detect 50\% of our target sources. The sources with lower K band luminosity at pc-scale suggest that synchrotron emission can still be detected in the high-frequency domain extending until or beyond 22 GHz without any interference from the thermal emission of dust and gas.

    \item We found 7 GPS sources ($\sim 60\%$) with parsec to sub-parsec scale central radio-emitting region ($\leq$ 1 pc), two sources with extended outflow-like structure (NGC 1052, NGC 4151), and two with multicomponent features (NGC 3079, NGC 5506). The turnover frequencies or peaks of the convex spectral shape range from $\sim$3-6 GHz. Five sources showed nearly flat or inverted spectra. A component of NGC 5506 (component B1) separated by 4--5 pc from the central source showed a steep spectrum.

    \item For the GPS sources, we found a linear correlation between turnover frequencies and lower frequency spectral indices or $\alpha_{1}$. This correlation means that the radio flux density drops more rapidly or gets absorbed (mainly synchrotron self-absorption in this range of frequencies) at lower frequencies for the sources with higher turnover frequencies. We established that most of our sources are mainly a group of young AGNs with dense central regions and relatively higher accretion rates. In addition, sources with flat spectra showed mostly resolved/slightly resolved bright central regions establishing that the radio emission is from pc-scale winds/shocks (interaction with ISM or star formation). On the other hand, sources with steep spectra may produce radio emission from a small-scale optically-thin, one-sided, strong collimated outflow. 

    \item Higher accretion rate favors wind-based radio emission in AGNs, similar to different accretion states observed in XRBs transitioning from the ``low/hard'' state to the ``high/soft" state where the radio emission is suppressed. We found that sources with flat/inverted spectra are grouped in the region of highest accretion and lowest radio luminosity, supporting the idea of wind-based radio emission or interaction between outflows and the neighboring ISM in the central region.

    \item A thermal free-free emission from an optically thick region in the immediate vicinity of a black hole may be the primary source of radio emission for the sources with flat/inverted spectra. The lower range of brightness temperature, especially in the K band or higher frequency regime, ranges between $10^{6.2}$~K$~\geq T_b\geq10^{5.2}$~K, suggesting compact central radio emission possibly harboring non-relativistic star-forming or thermal radio emitting components. 
\end{enumerate}}

VLBA high-resolution and multi-wavelength observations at radio energies allowed us to provide insights into a critical piece of the overall emission mechanisms of these AGNs. Our next goal is a high-resolution K band monitoring project of a large sample of nearby AGN, which can be crucial to identify the true central origin (and nature) of radio emission (TCORE). A future study using an even more powerful radio telescope, such as the higher-frequency Event Horizon Telescope (EHT) or Atacama Large Millimeter Array (ALMA), would be useful to constrain the particle acceleration mechanisms in the near-horizon environments of the supermassive black holes in galaxy centers.


\begin{acknowledgments}
This work supports USNO's ongoing research into the celestial reference frame and geodesy.\\

The National Radio Astronomy Observatory is a facility of the National Science Foundation operated under a cooperative agreement by Associated Universities, Inc. The authors acknowledge the use of the Very Long Baseline Array under the US Naval Observatory's time allocation. This work made use of data supplied by the UK Swift Science Data Centre at the University of Leicester. 
\end{acknowledgments}
%

\vspace{5mm}
\facilities{VLBA, Swift}

\software{\textsc{aips}, Astropy~\citep{Astropy_2013A&A...558A..33A,Astropy_2018AJ....156..123A}, \textsc{casa}~\citep{CASA_2022PASP..134k4501C}
          }




\bibliography{multiband}{}
\bibliographystyle{aasjournal}
\appendix 
\section{Additional Plots} \label{appendix: a}
\setcounter{figure}{0}
\renewcommand{\thefigure}{\Alph{section}\arabic{figure}}
\renewcommand*{\theHfigure}{\thefigure}
We plotted radio loudness vs. Eddington ratio, using the radio loudness parameter as ~$R_\mathrm{X} = L_\mathrm{6 GHz} / L_\mathrm{2-10 keV}~\&~L_\mathrm{8.6 GHz} / L_\mathrm{2-10 keV}$, where 6 GHz luminosities were calculated using the peak intensities (Paper I and III) and 8.6 GHz luminosities were measured from the total radio flux densities (5$\sigma$) of our observations (this work).

\begin{figure*}[hbt!]
\centering
\includegraphics[scale = 0.3]{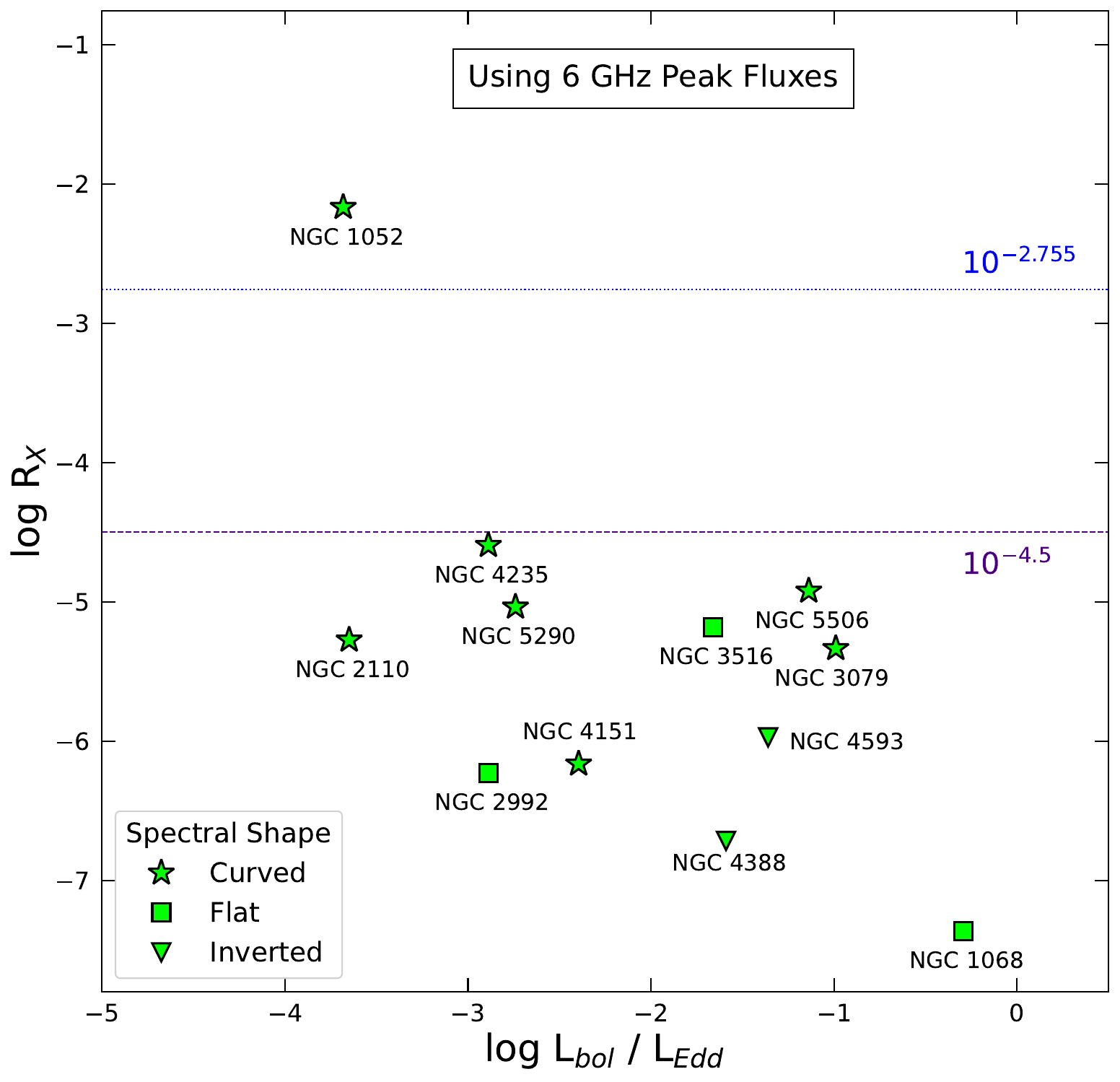}
\hspace{2mm}
\includegraphics[scale = 0.3]{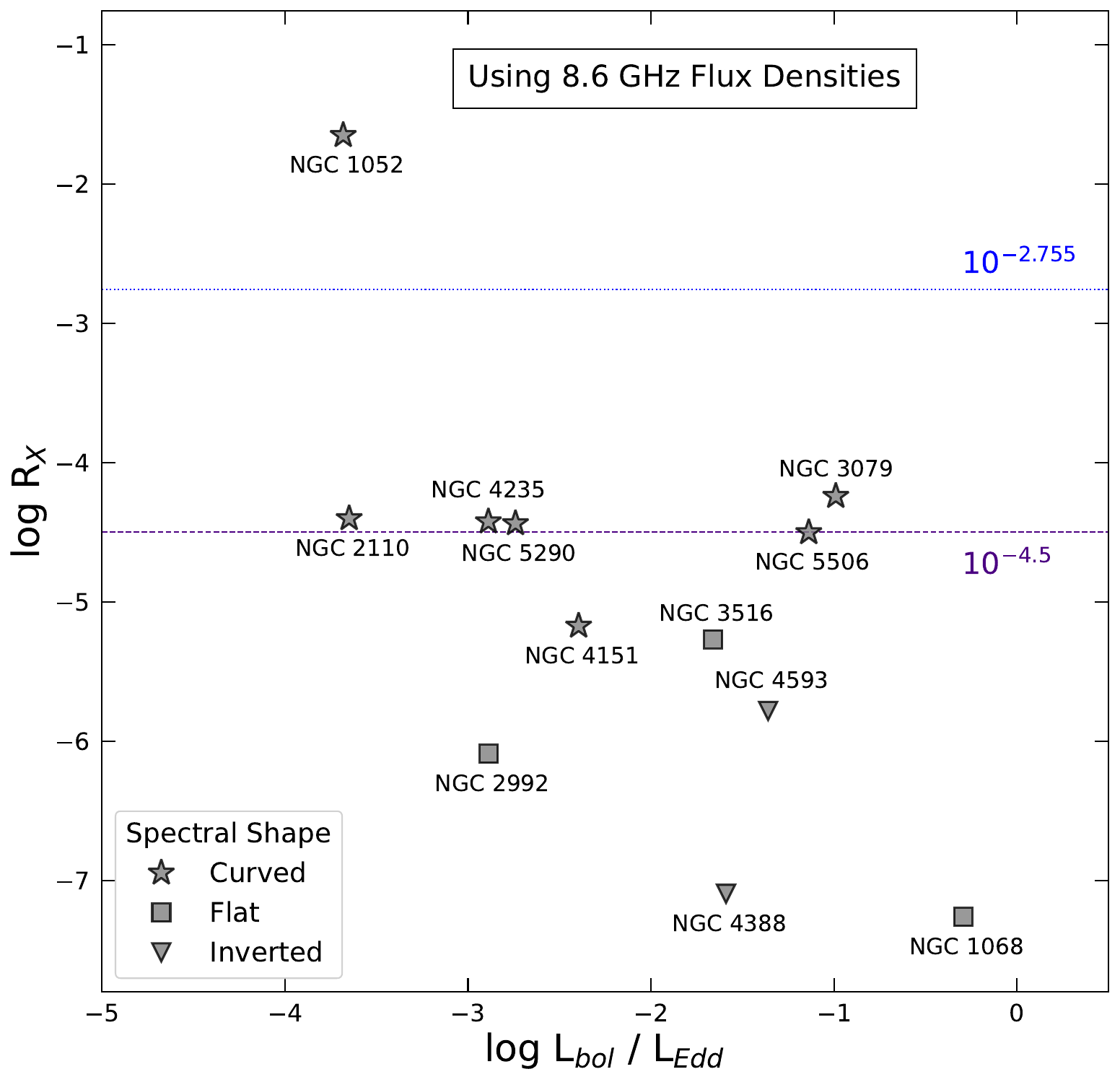}
\caption{Same as Figure~\ref{fig:Rx_vs_accretion}, but with 6 GHz peak intensities (\textit{left}) and 8.6 GHz flux densities  from the 5$\sigma$ contour region (\textit{right}).} 
\label{fig:Rx_vs_accretion_appendix}
\end{figure*}



\end{document}